\documentclass[twocolumn]{aa} 
\usepackage{epsfig}
\usepackage{amsmath}
\usepackage{amssymb}

\usepackage{natbib}
\usepackage{color}
\usepackage{relsize}

\usepackage{txfonts}
\begin{document}

   \title{Radiation Magnetohydrodynamics In Global Simulations Of Protoplanetary Disks.}

   \author{M. Flock
          \inst{1,2}
          \and
          S. Fromang\inst{1,2}
	  \and 
	  M. Gonz\'alez\inst{2,}\inst{3}
	  \and
	  B. Commer\c{c}on\inst{4}	  
          }

   \institute{CEA, Irfu, SAp, Centre de Saclay, 91191 Gif-sur-Yvette, France\\
     \email{Mario.Flock@cea.fr}
     \and 
     UMR AIM, CEA-CNRS-Univ. Paris Diderot, Centre de Saclay, 91191 Gif-sur-Yvette, France
     \and
     Université Paris Diderot, Sorbonne Paris Cité, AIM, UMR 7158, CEA, CNRS, F-91191 Gif-sur-Yvette, France
          \and
	     Laboratoire de radioastronomie, UMR 8112 du CNRS, \'{E}cole normale sup\'{e}rieure et Observatoire de Paris, 24 rue Lhomond, F-75231 Paris Cedex 05, France   
             }

   \date{}

  \abstract
   {}
   {Our aim is to study the thermal and dynamical evolution of
     protoplanetary disks in global simulations, including the physics
     of radiation transfer and magneto-hydrodynamic turbulence caused by the
     magneto--rotational instability.}  
   {We develop a radiative transfer method based on the flux-limited
     diffusion approximation that includes frequency dependent irradiation by
     the central star. This hybrid scheme is implemented
     in the PLUTO code.
     The focus of our implementation is on the performance
     of the radiative transfer method. Using an optimized Jacobi
     preconditioned BiCGSTAB solver, the radiative module is three
     times faster than the magneto--hydrodynamic step for the disk setup we
     consider. We obtain weak scaling efficiencies of $70\%$ up to $1024$ cores.}
   {We present the first global 3D radiation magneto-hydrodynamic simulations of a stratified protoplanetary disk. The disk model parameters are chosen to
     approximate those of the system AS 209 in the star-forming region
     Ophiuchus. Starting the simulation from a disk in radiative and
     hydrostatic equilibrium, the magneto--rotational instability quickly causes magneto--hydrodynamic turbulence and
     heating in the disk. We find that the turbulent properties are
     similar to that of recent locally isothermal global simulations of
     protoplanetary disks. For
     example, the rate of angular momentum transport $\alpha$ is a few times $10^{-3}$. For the disk parameters
     we use, turbulent dissipation heats the disk midplane and raises the 
     temperature by about $15 \%$ compared to passive disk models. 
     The vertical temperature profile shows no temperature peak at
     the midplane as in classical viscous disk models.  
     A roughly flat vertical temperature profile establishes in the disk
     optically thick region close to the midplane.
     We reproduce the vertical temperature profile with a viscous disk models for which the stress tensor 
     vertical profile is flat in the bulk of the disk and vanishes in
     the disk corona.} 
   {The present paper demonstrates for the first time that global
     radiation magneto--hydrodynamic simulations of turbulent protoplanetary disks are
     feasible with current computational facilities. This opens up the
     windows to a wide range of studies of the dynamics of
     protoplanetary disks inner parts, for which there are significant 
     observational constraints.} 


   \titlerunning{RMHD of protoplanetary disks}

   \maketitle
%

\section{Introduction}
The understanding of planet formation requires a deep insight into the
physics of protoplanetary disks.  
Recent observations of young disks in nearby star-forming regions
\citep{fur09,and09} have been able to constrain important physical
parameters, like the disk mass and radial extent, its flaring index or
the dust--to--gas mass ratio. Our understanding of these observations
is mainly based on 2D radiative viscous disk models
\citep{chi97,dal98,dul02b} that include proper dust opacities and
irradiation by the star. The energy released by the accretion process
is an important source for determining the structure and the evolution
of the inner disk regions. The magneto-rotational instability 
\citep[MRI,][]{bal98} is the most likely candidate to drive accretion
by an effective viscosity from magnetic turbulence. Up to now there is no
global model which combines both magneto-hydrodynamics (MHD) turbulence driven by the MRI and
the radiative transfer including irradiation by the star and proper
dust opacities. The main challenge to perform such simulations is the
computational effort. Global MHD simulations need high resolutions to
resolve the MRI properly \citep{fro06,flo10,sor12} and the
computational cost required to solve additional radiative transfer
equations remains a challenge. The first full radiation magneto-hydrodynamics (RMHD) of that
problem were
performed in local box simulations by \citet{tur03} using a flux-limited diffusion (FLD) approach \citep{lev81}. In the past few years, several
accretion disk simulations have been performed using similar numerical
schemes \citep{tur04,hir06,bla07,kro07,fla09} and 
recently including irradiation heating \citep{hir11}. More
sophisticated radiation hydrodynamics (RHD) methods, like the two-moment method
\citep{gon07} are usually very time demanding because they require large matrix inversion. 
In this work we develop a radiative transfer method based on the
two-temperature grey\footnote{A grey approach integrates over all frequencies} FLD approach by \citet{com11} and including frequency
dependent irradiation by the star \citep{kui10}. This
hybrid scheme captures accurately the irradiation energy by the
star and performs well compared to computational expensive Monte-Carlo
radiative transfer methods \citep{kui13}. 
We particularly focus on the serial and parallel performance of our method. The model we design is especially suited for global RMHD disk calculations.
Our paper is split into the following parts.
In section 2 we describe the RMHD equations, the numerical scheme and a performance test. 
In section 3 we explain our initial conditions for global RMHD disk calculations, the iteration method for calculating the disks radiative hydrostatic equilibrium and the boundary conditions. 
In section 4 we present the results, followed by the discussion and the conclusion. 
In the Appendix we show details of the discretization of the FLD method, the numerical developments and tests.   
\section{Numerical implementations}

\subsection{Equations and numerical scheme}
%
%
%
%

In this paper we solve the ideal RMHD equations using the FLD approximation and including irradiation by a central star. We use a spherical coordinate system $(r,\theta,\phi)$ which has advantages for the
treatment of stellar irradiation by means of a simple ray-tracing
approach and because it is well adapted to the flared structure of
protoplanetary disks. The set of equations reads  
\begin{eqnarray}
\rm \frac{\partial \rho}{\partial t} + \nabla \cdot \left [\rho \vec{v}\right ] &=&
0 \, , \label{eq:MDH_RHO} \\
\rm \frac{\partial \rho \vec{v}}{\partial t} + \nabla \cdot \left [
  \rho \vec{v} \vec{v}^T - \vec{B}\vec{B}^T \right ] + \rm \nabla P_t
&=& - \rho \nabla \Phi \, , \label{eq:MDH_MOM} \\
\rm \frac{\partial E}{\partial t} + \nabla \cdot \left [ (\rm E +
  P_t)\vec{v} - (\vec{v}\cdot\vec{B})\vec{B} \right ]  &=& - \rho
\vec{v} \cdot \nabla \Phi \nonumber\\& &  - \kappa_\mathrm{P}(\mathrm{T}) \rho c (\rm a_R T^4 - E_R
)\nonumber\\ & & - \nabla \cdot \mathrm{F}_* \, , \label{eq:MDH_EN} \\
\rm \partial_t E_R - \nabla \frac{c \lambda}{\kappa_\mathrm{R}(\mathrm{T}) \rho} \nabla
\rm E_R &=& + \kappa_\mathrm{P}(\mathrm{T}) \rho \mathrm{c} (\rm a_R T^4 -E_R) \, , \label{eq:ER} \\
\rm \frac{\partial B}{\partial t} + \nabla \times (\vec{v} \times
\vec{B}) &=& 0 \, , \label{eq:MDH_MAG} 
\end{eqnarray}
in which the two coupled equations for the radiation transfer are
\begin{eqnarray}
\rm \partial_t \rho \epsilon &=& - \kappa_\mathrm{P}(\mathrm{T}) \rho \mathrm{c} (\rm a_R T^4 - E_R) - \nabla \cdot F_* \, , \label{eq:RAD1}\\
\rm \partial_t E_R - \nabla \frac{c \lambda}{\kappa_\mathrm{R}(\mathrm{T}) \rho} \nabla
E_R &=& + \kappa_\mathrm{P}(\mathrm{T}) \rho \mathrm{c} (\rm a_R T^4 -E_R) \, , \label{eq:RAD2}
\end{eqnarray}
with the density $\rho$, the velocity vector $\vec{v}$, the magnetic
field vector\footnote{The magnetic field is normalized over
  the factor $1/\sqrt{4\pi}$} $\vec{B}$, the total pressure $\rm P_t = P +
0.5\vec{B}^2$, the gas pressure $\rm P= \rm \rho k_B T / (\mu_g u)$ with
the gas temperature $\rm T$, the mean molecular weight $\rm \mu_g$,
the Boltzmann constant $\rm k_B$, the atomic mass unit $\rm u$, the
gravitational potential $\rm \Phi = GM_*/r$ with the gravitational constant G, stellar mass $\rm M_*$, r the radial distance to the star, the total energy $\rm
E=\rho \epsilon + 0.5 \rho \vec{v}^2 + 0.5 \vec{B}^2$ with the
gas internal energy $\rho \epsilon$, the radiation energy E$_\mathrm{R}$, the irradiation flux $\mathrm{F}_*$, the Rosseland and
Planck mean opacity $\rm \kappa_R$ and $\rm \kappa_P$, the radiation constant $\rm a_R=(4 \sigma)/c$ with the Stefan-Boltzmann constant $\rm \sigma = 5.6704 \times 10^{-5}$ erg.cm$^{-2}$s$^{-1}$K$^{-4}$, and c the speed of light. To enforce causality, we use the flux limiter $\rm \lambda = (2 + R)/(6 + 3R + R^2)$ by \citet[][ Eq.~28 therein]{lev81}
with $\rm R = |\nabla E_R |/(\kappa_R \rho E_R)$. The closure
relation between gas pressure and internal energy 
is provided by the ideal gas equation of state $\mathrm{P}=(\Gamma -1) \rho
\epsilon$, with the adiabatic index $\Gamma$. We choose a mixture of hydrogen and helium with solar abundance \citep{dec78,bit13b} so that $\rm \mu_g =2.35$ and $\Gamma=1.42$.

After the MHD step, the method solves the two-coupled radiative transfer equations (\ref{eq:RAD1} and \ref{eq:RAD2}). We neglect in the equations all terms of the order $\rm v/c$ \citep{kru07}, including the radiation pressure terms and the radiation force in the momentum equations since $\rm E_R << \rho \epsilon$. 
These approximations are well suited for our applications ($\rm v \ll c$) but not necessarily for other regimes, like the dynamic diffusion \citep{kru07}.
A similar method was presented in \citet{bit13a} or \citet{kol13}.

It has recently been shown that frequency dependent irradiation is
more accurate in the context of protoplanetary disks to capture 
irradiation heating \citep{kui10,kui13}. The irradiation flux $\mathrm{F}_*$ at
a radius $\mathrm{r}$ is calculated as
\begin{equation}
\rm F_*(r) = \int_\Omega \int_\nu B_\nu (\nu,T_*) \left(\frac{R_*}{r} \right)^2 e^{-\tau(\nu,r)} \Omega d\nu d\Omega, 
\label{eq:IRRAD}
\end{equation}
with the Planck function $\rm B_\nu (\nu,\mathrm{T}_*)$, 
the solid angle $\Omega$, the surface temperature of the star $\rm T_*$, the radius of the star $\rm R_*$, the frequency $\nu$, and the radial optical depth for the irradiation flux
\begin{equation}
\rm \tau(\nu,r)=\int_{R_*}^r \kappa(\nu) \rho dr.
\label{eq:TAU}
\end{equation} 
The irradiation by the star is used as a source term in Eq.~3. This approximation is valid due to short penetration time for stellar rays through the domain compared to the longer hydrodynamical timescale.

\begin{figure}
\psfig{figure=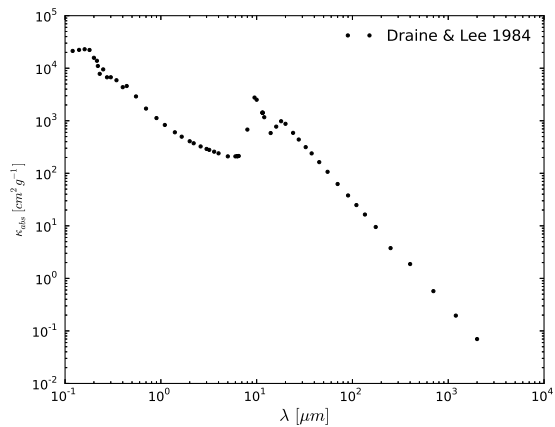,scale=0.40}
\caption{Dust absorption opacity over wavelengths. }
\label{fig:OPAC}
\end{figure}

Fig.~\ref{fig:OPAC} shows the frequency dependent dust absorption
opacity. The opacity tables are derived for
particle sizes of 1~$\mu m$ and below \citep{dra84}. 
We note that for the setup presented in this paper the temperature
stays below the dust evaporation temperature of about 1000 K so that we can neglect the gas opacities. 
To calculate the opacity involved in the RMHD equations we take into account the dust--to--gas mass ratio for small particles which we define as 1\% of the total dust--to--gas mass ratio $\rm \epsilon_{\mathrm{d2g}}$ \citep{bir12}. 
 
The ideal MHD equations are solved using the PLUTO code \citep{mig09}.
The PLUTO code is a highly modular, multidimensional and multi-geometry code that can be applied to relativistic or non-relativistic (magneto-)hydrodynamics flows. For this work we choose the Godunov type finite volume configuration which consists of a second order space reconstruction, a second order Runge-Kutta time integration, the constrained transport (CT) method \citep{gar05}, the orbital advection scheme FARGO MHD \citep{mas00,mig12}, the HLLD Riemann solver \citep{miy05}, and a Courant number of $0.3$. In this work we neglect the magnetic dissipation which would appear at the right hand side of Eq. 5. The effect of magnetic dissipation is discussed in the conclusion section. 
 
To solve Eq.~(\ref{eq:RAD1} and \ref{eq:RAD2}) we use an implicit method 
as the gas velocities are small compared to the speed of light. 
The implicit discretized equations in spherical coordinates can be found in the Appendix.
We rewrite the radiative transfer equations in the matrix form $\rm
Ax=b$ where x is the solution vector. The solution of this system involves a matrix inversion and is solved by an iterative method to minimize the residual $\rm r=Ax-b$ until a given accuracy is reached. We use the Jacobi-preconditioned BiCGSTAB solver based on the work by \citet{van92}. 
As a convergence criteria we use the reduction of the $\rm L_2$ norm of the
residual, $\rm ||r||_2/||r_{init}||_2 < 10^{-4}$. 

%
%

\subsection{Radiative transfer method validation}

\begin{figure}
\psfig{figure=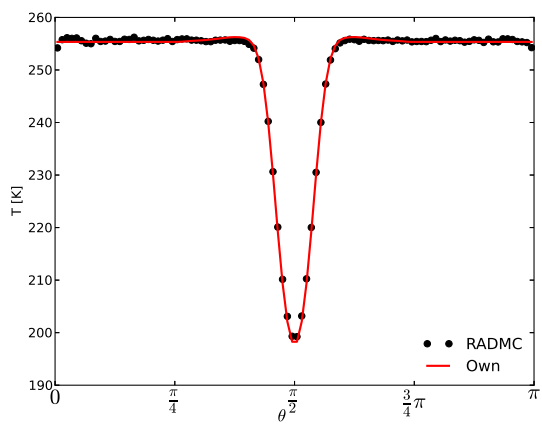,scale=0.40}
\psfig{figure=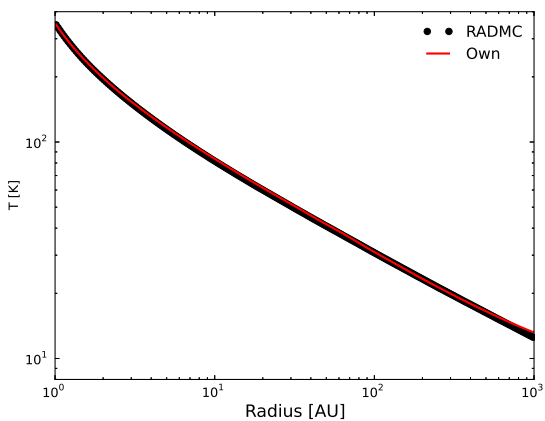,scale=0.40}
\caption{Left: $\theta$-temperature profile at a radius of 2~AU. Right: Radial temperature profile at the midplane.}
\label{fig:PASC}
\end{figure}
As a validation of our algorithm, we perform the radiation transfer
test for disks described by \citet{pas04}. We compute the
equilibrium temperature spatial distribution of a static disk irradiated by a star using different
radiative transfer methods. We compare
our method results with the one obtained using the Monte-Carlo radiative
transfer code RADMC-3D\footnote{www.ita.uni-heidelberg.de/\~{}dullemond/software/radmc-3d} \citep{dul12}. 
Following \citet{pas04}, we use the opacity table of \citet{dra84}, a dust--to--gas mass ratio of $0.01$, and frequency dependent irradiation with $61$ frequency bins. The
star parameters are $\rm T_* = 5800 \, K$ with $\rm R_* = 1 \, R_{\sun}$ and $\rm M_* =
\, 1 M_{\sun}$. The gas density follows  
\begin{equation}
\rm  \rho(r,z) = \rho_0 \left ( \frac{500\, AU}{r} \right ) \exp \left(
- \frac{\pi}{4} \left ( \frac{z}{h(r)} \right )^{2} \right ) \, ,
\label{eq:RHO_PASC}
\end{equation}
with 
\begin{equation}
\rm h(r)=125\, AU \left ( \frac{r}{500\, AU} \right )^{1.125} \, . 
\end{equation}
We present here the most optically thick disk configuration of \citet{pas04} for which $\rm \rho_0
= 8.321 \times 10^{-18}$  g.cm$^{-3}$. The initial temperature is 
set to $10$~K. The domain size ranges from $1$ to $1000$ AU in radius
and from $0$ to $\pi$ in $\theta$. We use a grid with logarithmically 
increasing cell size in radius and uniform in $\theta$. The overall
grid contains $240 \times 100$ cells. The boundary conditions of the
radiation energy are fixed to $10$~K in the poloidal direction and zero
gradient in the radial direction. We solve the radiative transfer
equations with a fixed time-step until we reach thermal equilibrium. 
The convergence criterion is $|res|_2/|res^{init}|_2 < 10^{-8}$.
In RADMC-3D we use $2.1\times 10^{10}$ photons. We plot the temperature
profile over radius and height in Fig.~\ref{fig:PASC}. Both profiles
agree very well with the results by RADMC-3D. In the
Appendix~\ref{test_sec}, we describe a resolution study of this test to determine the order of the scheme. 
We also perform an additional diffusion test for our FLD
method that we present for completeness. 

\subsection{Serial and parallel performance}

\begin{figure}
\psfig{figure=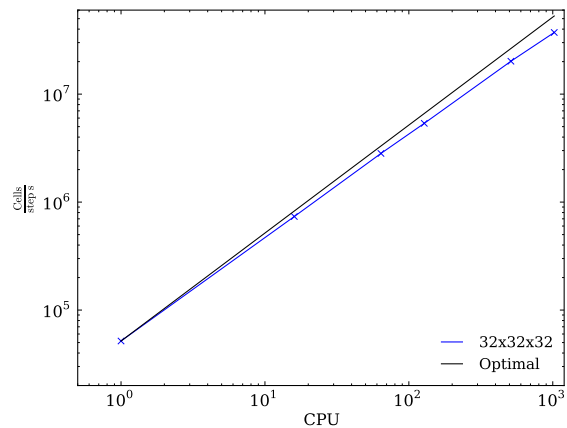,scale=0.40}
\caption{Weak scaling of the RMHD method ({\it blue line}) compared with
  optimal scaling ({\it black line}). Units are given in total number of
  grid cells per step per second.} 
\label{fig:BENCH}
\end{figure}

An important emphasis of our work was to develop a module with high
computational efficiency. During the development we focused on
reducing the number of operations per iteration cycle of the implicit
method. By increasing the memory usage we improved
substantially the performances of the algorithm. We performed an analysis of our module for the fiducial disk model (see
section~\ref{results_sec}). The MHD part takes $76.3 \%$ while the
radiation module consumes $22.5 \%$ of a full RMHD step. In our module
each matrix vector multiplications per iteration in the BiCGSTAB method have the
main computational cost with $8.2 \%$ from the full RMHD step,
followed by the frequency dependent irradiation with $5.9 \%$. 
%
%
In the following we present a weak scaling test. The number of iterations by
the matrix solver is fixed to 30 which is a typical value for reaching convergence in our
global high resolution model. We use $32^3$ cells per cpu as in this model, for
which we use $1024$ cores. Fig.~\ref{fig:BENCH} shows the parallel
performance of the RMHD module using an Intel Xeon 2.27 Ghz
system. With $1024$ cores we reach a scaling efficiency of 70.3$\%$ which is acceptable given the non-local nature of the
algorithm.
Nevertheless, the parallel performance is lower than the pure Godunov
scheme \citep{mig07}. This is largely due to
the important amount of communications inside the BiCGSTAB method. Here one has to distribute three times all
neighboring cells for each core per iteration to compute the new residual. As the number of
iterations strongly depends on the physical problem, we do not a priori
know the serial and the parallel performance for a given application
beforehand. We
note that the parallel performance depends also on the ratio
between the number of grid cells to communicate over the grid cells
per core. Reducing the number of cells below $32^3$ per core would result in 
reduced scaling performances.

\section{Initial disk structure and boundary conditions}

As an illustration of the possibilities offered by our numerical
implementation of a FLD scheme within the PLUTO code, we present in
the remaining of this paper a series of 3D RMHD simulations of
a fully turbulent protoplanetary disk. In this section, we describe
the disk model parameters and the iterative procedure we design to
construct an initial setup that is in hydrostatic
equilibrium. Boundary conditions in these simulations turned out to
be subtle, so we also detail the set of conditions we used in this 
particular case. We caution the reader that finding a proper set of
boundary conditions is delicate and likely to be problem dependent.

\subsection{An iterative procedure}
\label{iterative_sec}

Finding an irradiated disk structure in hydrostatic equilibrium is not
a straightforward task. This is because, for a given irradiation
source, the disk temperature depends on the spatial density
distribution which itself depends on temperature (because of the
pressure force). We thus solve for the hydrostatic disk structure
iteratively: assuming a given density in the disk, we calculate
the temperature as a result of disk irradiation using the hybrid FLD module
described above. To calculate the new radiation and temperature field we use the typical diffusion time for this problem. 
The diffusion time can be estimated by $\rm \Delta t_{dif} \sim 3 \kappa \rho \Delta x^2/c $. Using typical values at 1~AU with solar parameters $\rho=10^{-10}$ g.cm$^{-3}$, $\kappa=1$ cm$^2$ g$^{-1}$ and $\Delta x=1$ AU, we obtain $\rm \Delta t_{dif}\sim 10^{6}$s. After the temperature and radiation field have reached equilibrium, a new density profile is calculated and the algorithm iterates until convergence. 

The following input parameters are needed: the surface density over
radius $\Sigma(\mathrm{r})$, the opacity including the dust-to-gas mass ratio
and the stellar parameters T$_*$, M$_*$, and R$_*$. The density and the
azimuthal velocity $\rm v_{\phi}$ are updated integrating the equations of
hydrostatic equilibrium in spherical coordinates using a second order
Runge Kutta method. In hydrostatic equilibrium, these are, for the
radial and poloidal direction respectively
\begin{eqnarray}
\rm \frac{\partial \mathrm{P}}{\partial \mathrm{r}} &=& - \rho \frac{\partial \Phi}{\partial
\rm  r} + \frac{\rho \rm v^2_\phi}{\rm r} \label{eq:P_R} \\
\rm \frac{1}{r} \frac{\partial \mathrm{P}}{\partial \theta} &=&
\rm \frac{1}{\tan{\theta}}\frac{\rho v^2_\phi}{r} \label{eq:P_T}.
\end{eqnarray}
%
%
%
Once $\rm \rho(r,\theta_0)$ is known for a given value of $\rm \theta_0$ (e.g. $\theta_0=\pi/2$ for the midplane) 
and for all $\rm r$, Eq.~(\ref{eq:P_R}) can be used to calculate
$\rm v_{\phi}(r,\theta_0)$. The second equation is then integrated to give the density field at the next interface $\theta_0+\Delta \theta/2$ for any value of $\rm r$ and we can repeat the cycle. Using the mid-point integral method we reach second order accuracy.
We impose the midplane gas density using
\begin{equation}
\rm \rho(r,\pi/2) = \frac{\Sigma(\mathrm{r})}{\sqrt{2\pi}\mathrm{H}} \, ,
\end{equation}
where $ \mathrm{H}/\mathrm{r}=\sqrt{\tilde{\mathrm{T}}\mathrm{r}/(G\mathrm{M}_*})$ and $\rm \tilde{T} =
(k_B T) /(\mu_g u)$. 
The above relation is only valid for a constant vertical temperature and so a Gaussian vertical density profile. 
We do not expect this to be
the case here since the temperature can a priori vary with distance to
the midplane. However, the bulk of the disk around the midplane has a
constant vertical temperature (see Fig.~\ref{fig:RMHD} and
section~\ref{as209_sec} below). Since this is the location where most
of the mass is located, the actual surface 
density is close to the targeted value with small deviations of
$10^{-3}$. To reach a higher accuracy, we multiply the density in each
grid cell by a constant factor $\rm \Sigma_{target}/\Sigma$ so that we
reach the target value before calculating the new temperature. We
iterate the procedure until both the temperature and density field
have converged to better than $\rm max((\mathrm{T}^{\mathrm{n}+1}-T^{n})/\mathrm{T}^\mathrm{n},(\rho^{\mathrm{n}+1}-\rho^{n})/\rho^{\mathrm{n}})
\le 10^{-8}$, where $\rm n$ is the number of iterations step. A validation
of this iterative method is presented in Appendix~\ref{hydrostatic_sec} by
reproducing the passive disk model of \citet{chi97}.

\subsection{Simulations parameters: the case of AS 209}
\label{as209_sec}

\begin{figure}
\hspace{-0.5cm}
\begin{minipage}{0.3\textwidth}
\psfig{figure=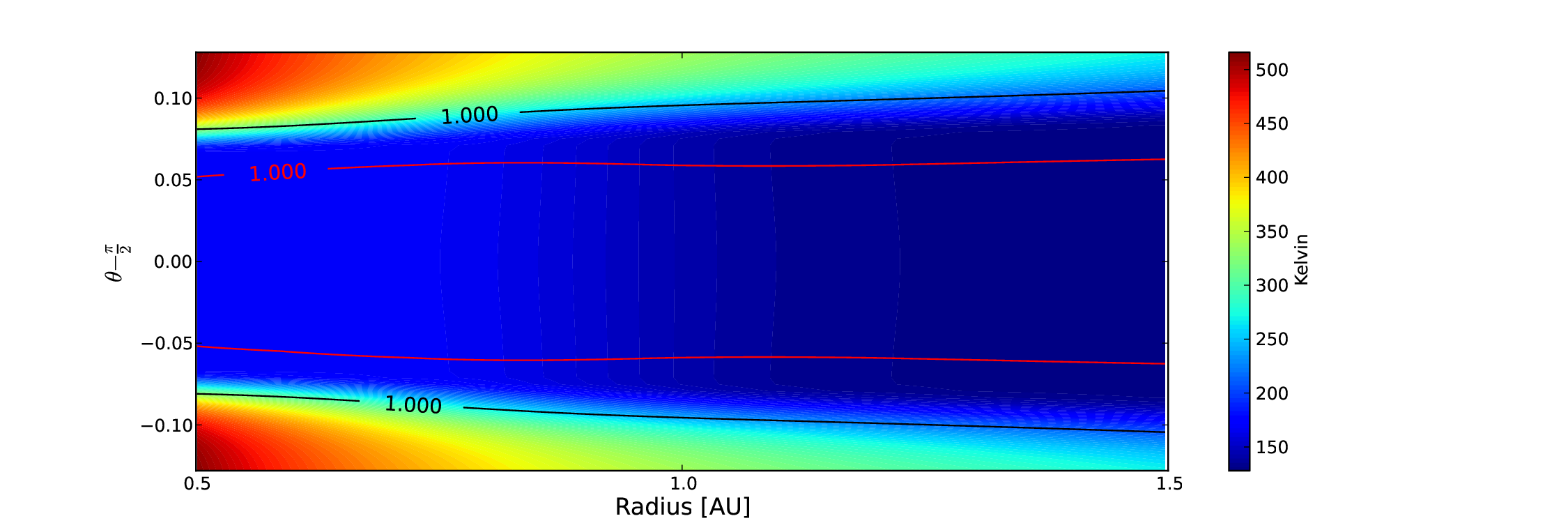,scale=0.30}
\psfig{figure=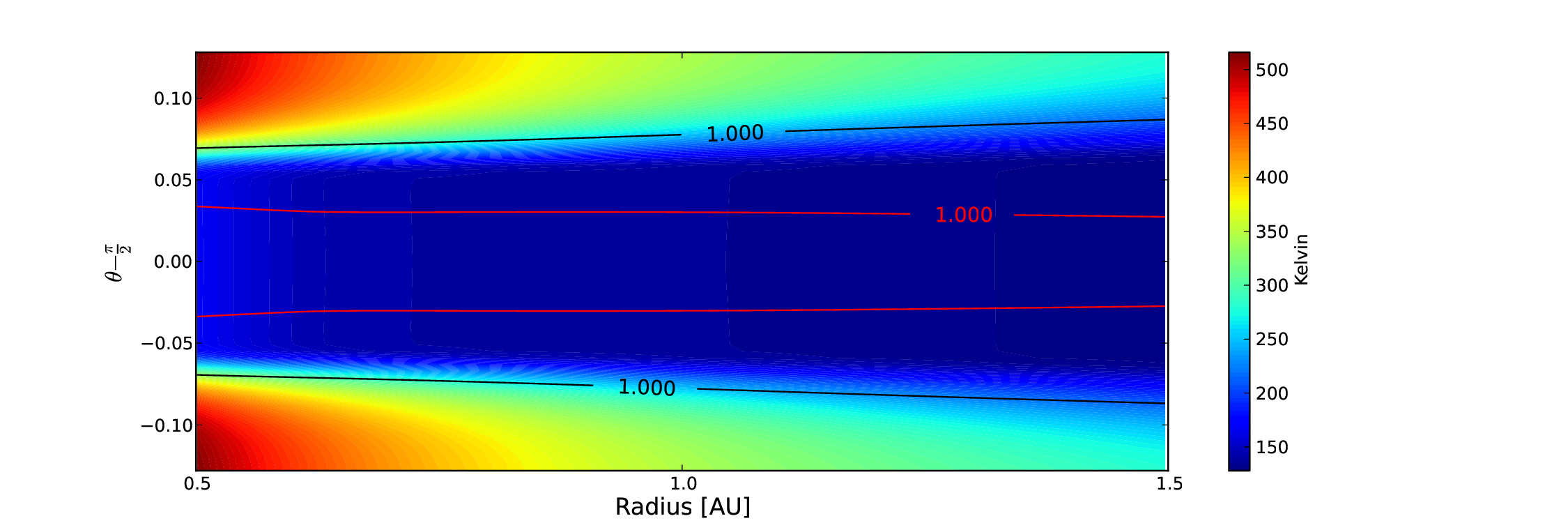,scale=0.30}
\psfig{figure=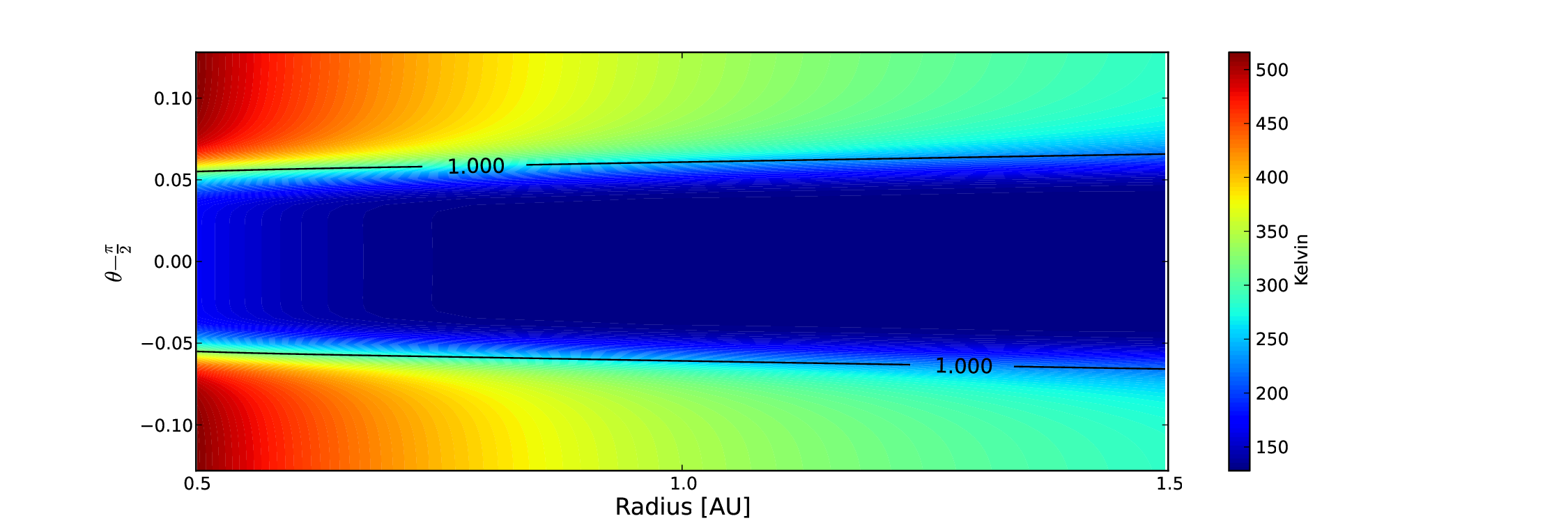,scale=0.30}
\end{minipage}
\caption{Initial temperature distribution for different amounts of small
  size dust ($\rm \le 1\mu m$), calculated from radiative hydrostatic
  equilibrium. From top to
  bottom: $\rm \Sigma_{dust}(1\, AU)= 0.17,0.017,0.0017$ g.cm$^{-2}$. We
  overplot the radial integrated $\tau=1$ line for the irradiation
  ({\it black solid line}) and the vertically integrated $\tau =1$ for the
  local thermal emission ({\it red solid line}).}  
\label{fig:RMHD}
\end{figure}

We use the scheme detailed above to calculate the initial disk
structure of the series of simulations we present in
section~\ref{results_sec}. We choose stellar and 
disk parameters inspired from those of the circumstellar disk AS 209
in the Ophiuchus star-forming region, for which there
are a number of observational constraints \citep{koe95,and09,per12}. 
Stellar mass, radius, and surface temperature are  well
  constrained parameters and we adopt the same values as \citet{and09}
  with 0.9 M$_{\sun}$, 2.3 R$_{\sun}$, and 4250 K. The gas surface 
density $\Sigma$ in the inner region of AS 209 is nearly a free
  parameter and only constrained indirectly by the dust surface
  density.  \citet{and09} estimated for this system a total dust
  surface density (including all particle sizes) of less than  $\rm 1$ g.cm$^{-2}$ at 1 AU from the star.  

For radiation hydrodynamics, the distribution of the small particles
($\le 1$~$\mu$m) is most important. This is because these 
particles dominate the opacity at optical, near- and mid-infrared wavelengths. They contain most of the 
dust surface, which is the controlling parameter for the opacity. 
By contrast, most of the dust mass, $\sim 99\%$ is stored in larger
particles \citep{bir12}. Roughly speaking, the dust surface density of
the small particles can be estimated to be less than $\rm 0.01$ g.cm$^{-2}$, using the parameters by \citet{and09}. We have calculated
our disk structure for three different amounts of small size dust 
particles: $\rm \Sigma_{dust}(1\, AU)= 0.17,0.017,0.0017$ g.cm$^{-2}$.
The initial temperature distribution in radiative
hydrostatic equilibrium for the three cases is shown in
Fig.~\ref{fig:RMHD}. In the top panel, $\rm \Sigma_{dust}(1\, AU)= 0.17$ g.cm$^{-2}$ and the disk
displays a large optical thick region. In the bottom panel, having $\rm \Sigma_{dust}(1\, AU)= 0.0017$ g.cm$^{-2}$, we find a very extended heated upper region while the disk midplane is completely optically thin to its own thermal
radiation. Between these two extrema we choose our fiducial model with a 
dust surface density of $\rm \Sigma_{dust}(1\, AU)= 0.017$ g.cm$^{-2}$. In this
configuration (shown on the middle panel of Fig.~\ref{fig:RMHD}),
both an optically thick midplane and an extended optically thin
corona fit within the computational domain.
The gas surface density is set to $\rm \Sigma(r)= 1700$ g.cm$^{-2} (\rm r/1\, AU)^{-0.9}$, such that
it equals the minimum mass solar nebula (MMSN) value at $1$~AU but
displays a shallower slope than the MMSN such as suggested by \citet{and09}, even if those constraints come from larger radial
distances. Assuming again that small particles carry only 1\% of the total dust mass,
we obtain a total dust--to--gas mass ratio $\rm \epsilon_{d2g}$ of $10^{-3}$ in our fiducial model.

The simulation spans the radial range $r=0.5-1.5$~AU, the poloidal range $\theta =\pi/2 \pm 0.13$, and the azimuthal range $\phi =0-\pi/3$. 
%
%
%
For the fiducial model in initial equilibrium, we obtain $\rm H/r
= 0.02$ at $\rm 0.5\, AU$ which results in $\pm 6.5$ scale heights
fitting in the computational domain at its inner boundary. Due to the
disk flaring the value of $\rm H/r$ at the outer radius is larger,
with $\rm H/r = 0.03$ corresponding to $\pm 4.3$ scale 
heights at the outer boundary. 

\subsection{Boundary conditions}
%
%
%
%
Finding suitable boundary conditions for a stable and physically
reasonable global simulation such as presented in this paper is
already difficult in ideal MHD. Radiative transfer makes the problem
even more difficult. In this section we describe in detail the
boundary conditions we have designed for that purpose. Straightforward
periodic boundary conditions are used for all variables in the
azimuthal direction, so we focus on the radial and poloidal boundaries.

In the radial direction we extrapolate linearly the density and azimuthal velocity. Radial and poloidal velocities are all set to
zero gradient. In the case of inflowing gas with a Mach
number of $0.1$ or higher, we force the radial velocity boundary condition to be reflective. The
poloidal and toroidal magnetic field components are set to follow a
$1/\mathrm{r}$ profile, while the radial magnetic field is calculated to ensure $\nabla \cdot \vec{B}=0$ in the ghost cells. Temperature is
set to zero gradient and we use $\rm E_{R}=a_R \mathrm{T}_0^4$ for the
radiation energy density $\rm E_{R}$ with $\mathrm{T}_0$ being the initial
radiative hydrostatic equilibrium temperature.
In order to avoid irradiation from directly illuminating the first
radial cell of the computational domain, we set a vertical dependent optical depth $\rm \tau_{init}(\theta)=\kappa_{0}(\theta) \rho_{0}(\theta) ({r_{0}-6\mathrm{R}_*})$ with the subscript 0 corresponding to the first cell in the radial direction. 
Using six stellar radii as the inner disk edge is a reasonable approximation to absorb
most of the irradiation at the disk midplane, as would be done by the
inner parts of the disk\footnote{Young stellar objects have dominant magnetic fields inside a few stellar radii, which destroy the disc structure \citep{gun13}}. Accordingly, the Rosseland opacity at the
radial boundary is modified in the optically thick region so that $\rm \kappa_{0}\rho_{0} \Delta
r < 0.1$. Even if this boundary condition seems unrealistic it will
prevent an artificial pile up of the radiation energy density near the boundary, and so permits the correct disk flaring (see Appendix~\ref{hydrostatic_sec}). 
 
In the poloidal direction we force the density to drop
exponentially. The velocities are set to zero gradient. In case of
inflowing poloidal velocities we reflect $\rm v_{\theta}$ in the ghost
cells. Tangential magnetic fields are set to zero gradient, while we
calculate the poloidal field so as to enforce $\rm \nabla \cdot
\vec{B}=0$. The temperature is set to zero gradient and the radiation
energy density is fixed to $\rm E_{R}=a_R T_{min}^4$ with $\rm T_{min}=10$ K. The temperature in the poloidal direction has to be small $\rm T_{min} << T_{gas}$ 
to ensure the disk can cool by radiating its energy away.

\section{Results}
\label{results_sec}

\begin{table*}[t]
\begin{center}
\begin{tabular}{cccccccccc}
\hline
\hline
Model & Resolution  & Radius:$\theta$:$\phi$ & Method & $\rm \epsilon_{d2g} $ & $\rm <T_{r\phi}> dyn.$ $\rm cm^{-2}$  & $\langle
\alpha \rangle$ & Orbits \\
\hline
\hline
L3D    & 256x64x256  & 0.5-1.5:$\pi/2 \pm$ 0.13:$0-\pi/3$ & RMHD & $10^{-3}$ & $0.037$  & $2.6 \cdot 10^{-3}$  & 0-650\\ 
$\rm L3D^{l}$  & 256x64x256  & 0.5-1.5:$\pi/2 \pm$ 0.13:$0-\pi/3$ & RMHD & $10^{-4}$ &  $0.062$  & $3.25 \cdot 10^{-3}$  & 0-650\\ 
\hline
H3D    & 512x128x512 & 0.5-1.5:$\pi/2 \pm$ 0.13:$0-\pi/3$ & RMHD & $10^{-3}$ & 0.076 & $4.6\cdot10^{-3}$ &  300-600\\ 
H3D-ISO& 512x128x512 & 0.5-1.5:$\pi/2 \pm$ 0.13:$0-\pi/3$ & MHD & $10^{-3}$ &  0.044  & $2.5\cdot10^{-3}$ & 300-600\\ 
\hline
H2D    & 512x128x1  & 0.5-1.5:$\pi/2 \pm$ 0.13:$-$ & RHD & $10^{-3}$ & - & $4.6\cdot10^{-3}$ & 0-200\\
$\rm H2D^{*}$    & 512x128x1  & 0.5-1.5:$\pi/2 \pm$ 0.13:$-$ & RHD & $10^{-3}$ & - & $4.6\cdot10^{-3}$  & 0-200\\
\hline
\end{tabular}
\caption{Model; Resolution; Domain ; Method ; Dust to gas mass ratio ;
  Total accretion stress in cgs units; Normalized accretion stress ;
  Inner orbits    } 
\label{tab:M3D}
\end{center}
\end{table*}

Table 1 summarizes the models we performed and provides an overview of
the integrated angular momentum transport properties we obtain. 
Models $\rm L3D$ and $\rm L3D^l$ are low resolution RMHD
simulations, with $\rm (N_r,N_{\theta},N_{\phi})=(256,64,256)$. Such a low
resolution enables long integration times of about 650 inner orbits. 
While the dust--to--gas mass ratio in model $\rm L3D$ is equal to
our fiducial value, $10^{-3}$, it is reduced by one order of magnitude
in model $\rm L3D^l$. To save computational time, we interpolate the
results of model $\rm L3D$ after 300 inner orbits on a grid twice as fine. 
After this time, the MRI has saturated, and we use the interpolated magnetic fields to restart the simulation, which constitutes model $\rm H3D$. This represents the fiducial model of the present paper and is
described in detail in the following sections. In Appendix~\ref{restart_sec} we describe 
the procedure of interpolation and restarting.  
To connect with previous work, its properties are compared 
with model H3D-ISO which is a locally isothermal model that uses
an azimuthal- and time-averaged temperature profile calculated from
the results of model $\rm H3D$ (see
section~\ref{varparams_sec}). Finally, we compare our results with a
couple of 2D radiative hydrodynamic simulations (performed in the disk
poloidal plane) of viscous disks, namely models $\rm H2D$ and $\rm
H2D^*$, that use different prescriptions for the viscous stress tensor
(see section~\ref{temperature_sec}). All MHD simulations are initialized
with a pure toroidal magnetic field with a uniform plasma beta value $\beta=2 \rm P/B^2=40$. Initial random 
velocity fluctuations are added to the initial disk configuration with
an amplitude equal to $10^{-3}$ of the local speed of sound. 


\subsection{Model H3D}

\subsubsection{Turbulent properties}
\label{turb_prop_sec}
\begin{figure*}
\psfig{figure=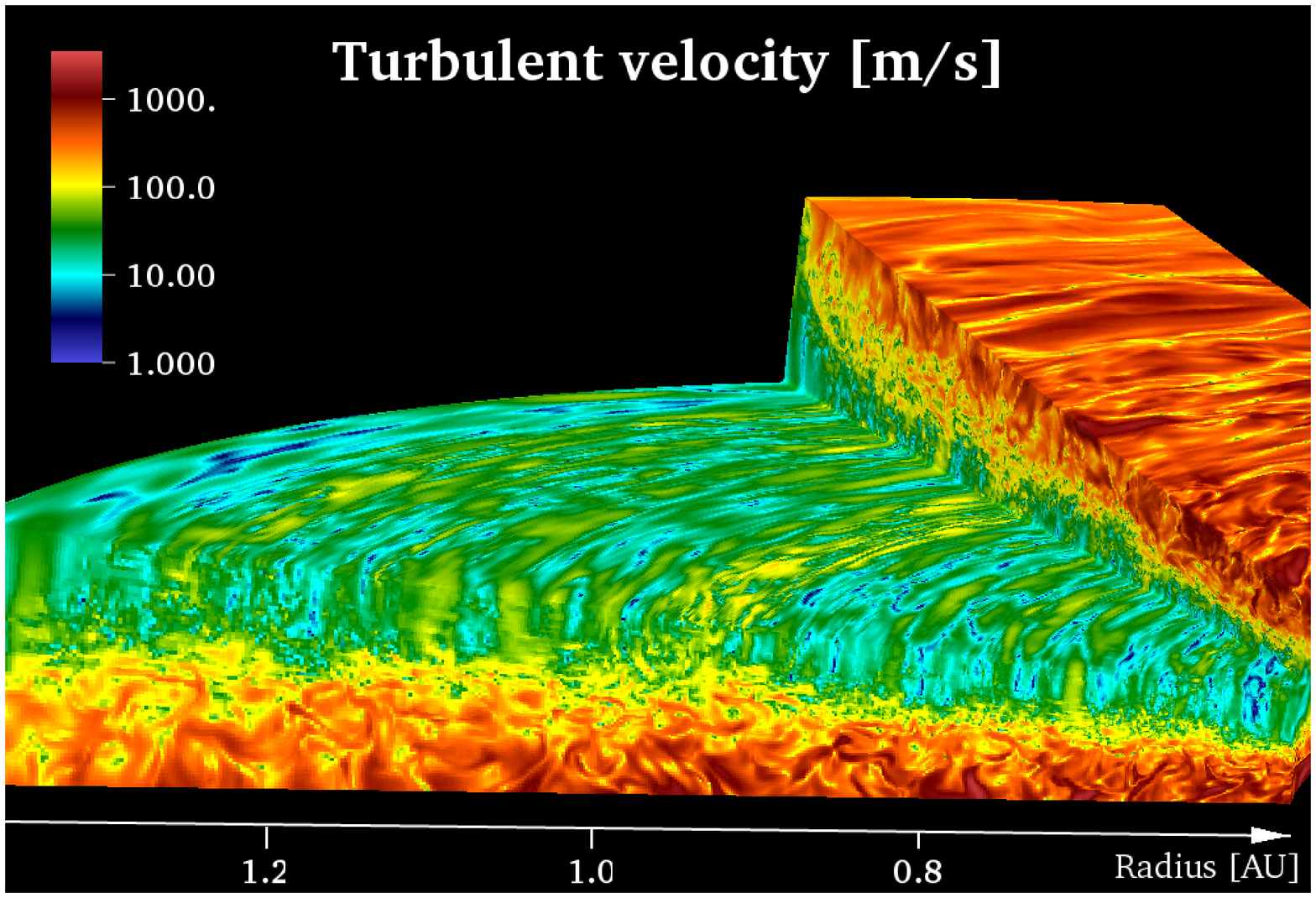,scale=0.4}
\psfig{figure=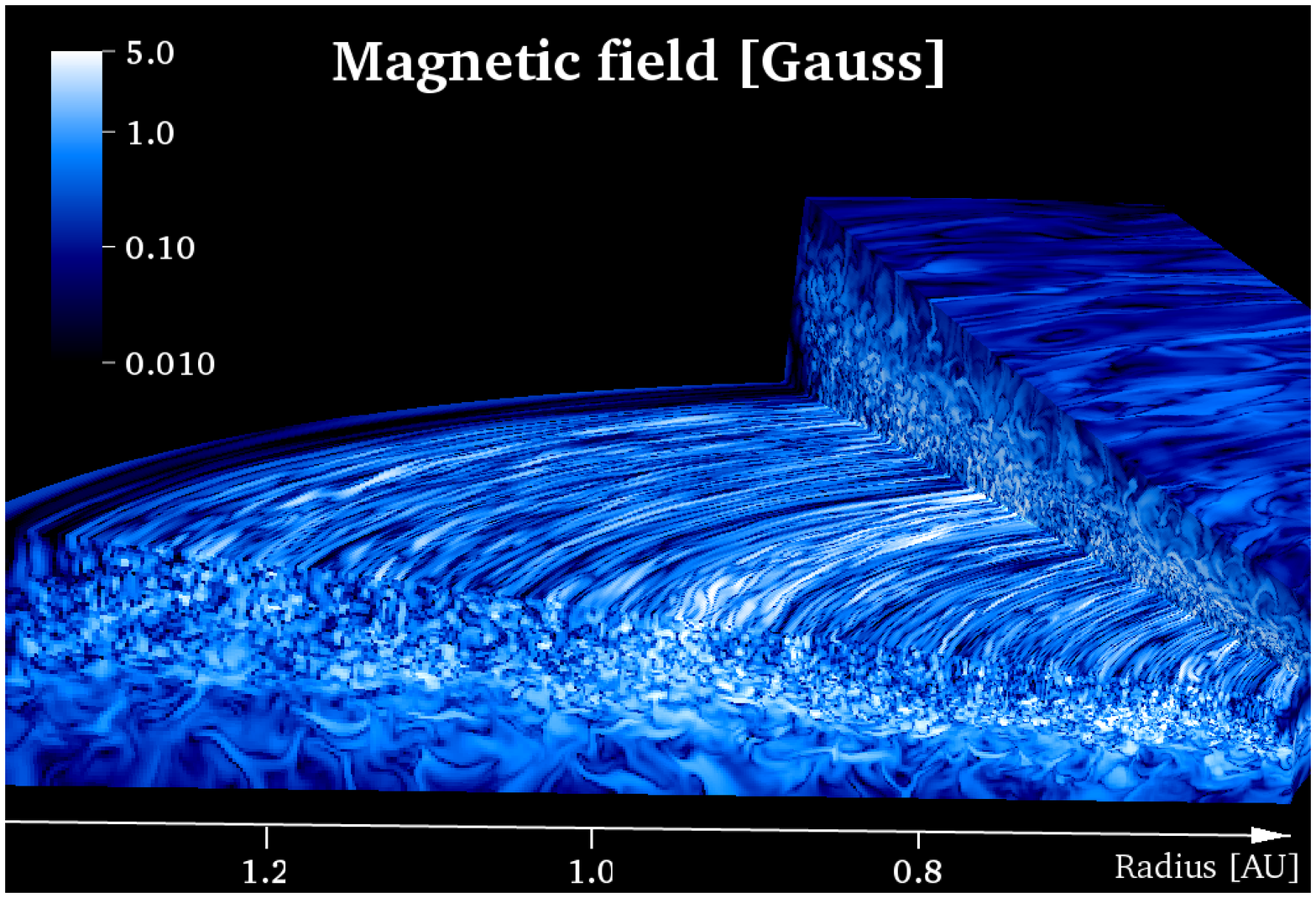,scale=0.4}
\caption{Snapshot of turbulent RMS velocity ({\it left panel}) and magnetic
  field strength ({\it right panel}) at the final time of the full RMHD simulation H3D.}  
\label{fig:VRMS-3D}
\end{figure*}
%
%
\begin{figure}
\psfig{figure=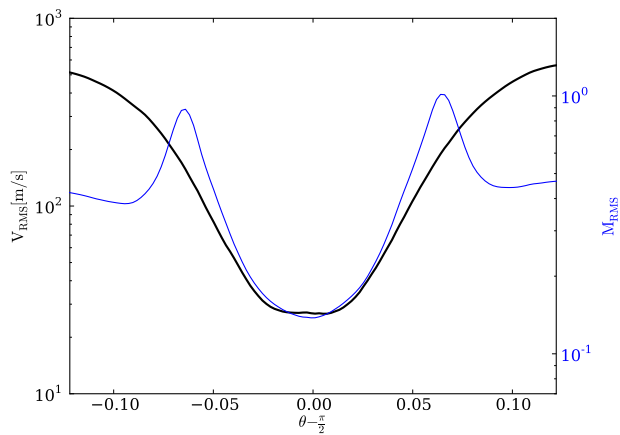,scale=0.40}
\caption{Vertical profile of turbulent RMS velocities for model H3D in units of m.s$^{-1}$ (black)
  and the corresponding turbulent Mach number (blue). Space
  average is done over azimuth at 1~AU with a time average of 100 inner
  orbits.}  
\label{fig:VRMS-L}
\end{figure}
\begin{figure}
\psfig{figure=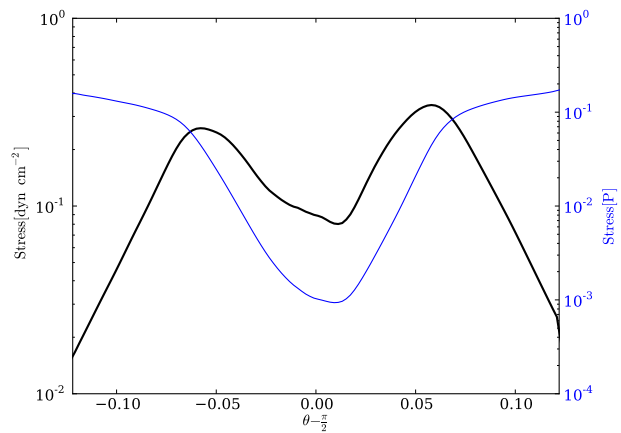,scale=0.40}
\caption{Vertical stress profile for model H3D in units of dyn.cm$^{-2}$ (black) and
  normalized using the local pressure (blue). Space
  average is done over azimuth at 1~AU with a time average of 100 inner
  orbits.} 
\label{fig:Stress-L}
\end{figure}
We start with a general description of the integrated properties of the 
turbulence in model $\rm H3D$. The turbulent nature of the flow is
best illustrated by Fig.~\ref{fig:VRMS-3D} which shows two snapshots of
the gas velocity fluctuations and the magnetic field strength
in the disk. The root--mean--squared (RMS) velocities in the disk
midplane range from 1 m.s$^{-1}$ up to 100 m.s$^{-1}$. In the corona the
turbulent velocities increase above 1000 m.s$^{-1}$. This is
consistent with the azimuthal and time averaged vertical profile of
the gas turbulent velocities as shown in Fig.~\ref{fig:VRMS-L} ({\it
black curve}). When normalized by the local sound speed ({\it blue
curve}), the plot shows a local Mach number around $0.15$ at the
midplane. This is typical of values obtained in isothermal simulations
\citep{fro06,flo10}. However, in contrast with such simulations, the
turbulent Mach number reaches a peak (with roughly sonic velocity
fluctuations) at around $\theta-\pi/2=0.07$, which corresponds to
  around two pressure scale heights.  
The Mach number then decreases above the peak location as the
temperature (and so the sound speed) rises faster than the turbulent
velocity. 

The right panel of Fig.~\ref{fig:VRMS-3D} shows the magnetic field
strength spatial distribution. In the midplane, it ranges from $0.01$
Gauss up to $5$ Gauss. In the corona, the field shows larger and
smoother fluctuations with values below $1$ Gauss. 
That magnetic field is largely
responsible for angular momentum radial transport in the disk. In
Fig.~\ref{fig:Stress-L}, we plot the vertical profile of the accretion stress which is the sum of Reynolds and Maxwell stresses 
\begin{equation}
\rm <T_{r\phi}> = \left \langle  \rho v'_{\phi}v'_{r} -
B_{\phi}B_{r} \right \rangle,
\label{eq:TRP}
\end{equation}
where $\rm v'_{r,\phi}$ is the radial and azimuthal velocity fluctuations and $<.>$ represents the time average.
The black curve corresponds to the absolute values of the stress (in cgs units).
%
%
It displays a peak of $\rm \sim 0.3$ dyn.cm$^{-2}$
at about $\theta-\pi/2 \sim 0.06$, a plateau with a small drop close
to the midplane ($\rm \sim 0.1$ dyn.cm$^{-2}$), and decreases in the disk
corona. Such a profile is in qualitative agreement with results
of isothermal local simulations \citep{sim12}
as well as in radiative MHD local simulations
\citep{hir06,fla10}, and in locally isothermal global MHD simulation
\citep{fro06}. By normalizing the stress over the local pressure
({\it blue curve}), we also obtain a vertical profile and
absolute $\alpha$ values similar to isothermal simulations
\citep{flo11}. $\alpha$ amounts to about $10^{-3}$ in the disk equatorial
plane and increases up to a few times $10^{-1}$ in the corona. Table 1
further shows the value of the accretion stress volume averaged over
the entire computational domain. Following \citet{flo11}, its
normalized value $\langle \alpha \rangle$ is defined according to 
\begin{equation}
\rm <\alpha> = \left\langle \frac{ \int \rho \Bigg(
\frac{\rho v'_{\phi}v'_{r}}{P} -
\frac{B_{\phi}B_{r}}{P}\Bigg)dV} {\int \rho dV} \right\rangle.
\label{eq:ALPHA}
\end{equation}
Time average is done between 400 and 500 inner orbits. Again, we find
typical values of about a few times $10^{-3}$, very similar to transport
coefficients measured in locally isothermal global simulations of
turbulent protoplanetary disks in the last few years. We note that the
changes in the turbulence properties over radius remain small 
during the simulation. Nevertheless, there are regions of increased
activity. This is the case for example of the region that is about $0.1$
AU broad visible in the 3D snapshots close to $R \sim 1$~AU at the midplane
(Fig.~\ref{fig:VRMS-3D}), where turbulent velocities and magnetic fields are larger. Such zones of increased turbulent activity could be
connected to long-lived zonal flows such as observed in local
\citep{joh09,dit13} and global simulations \citep{dzy10,flo11} when
using an isothermal equation of state. We now move in the following to
the specificities of the present work that are associated with
radiative transfer.



\subsubsection{Temperature evolution}
\label{temperature_sec}

\begin{figure}
\psfig{figure=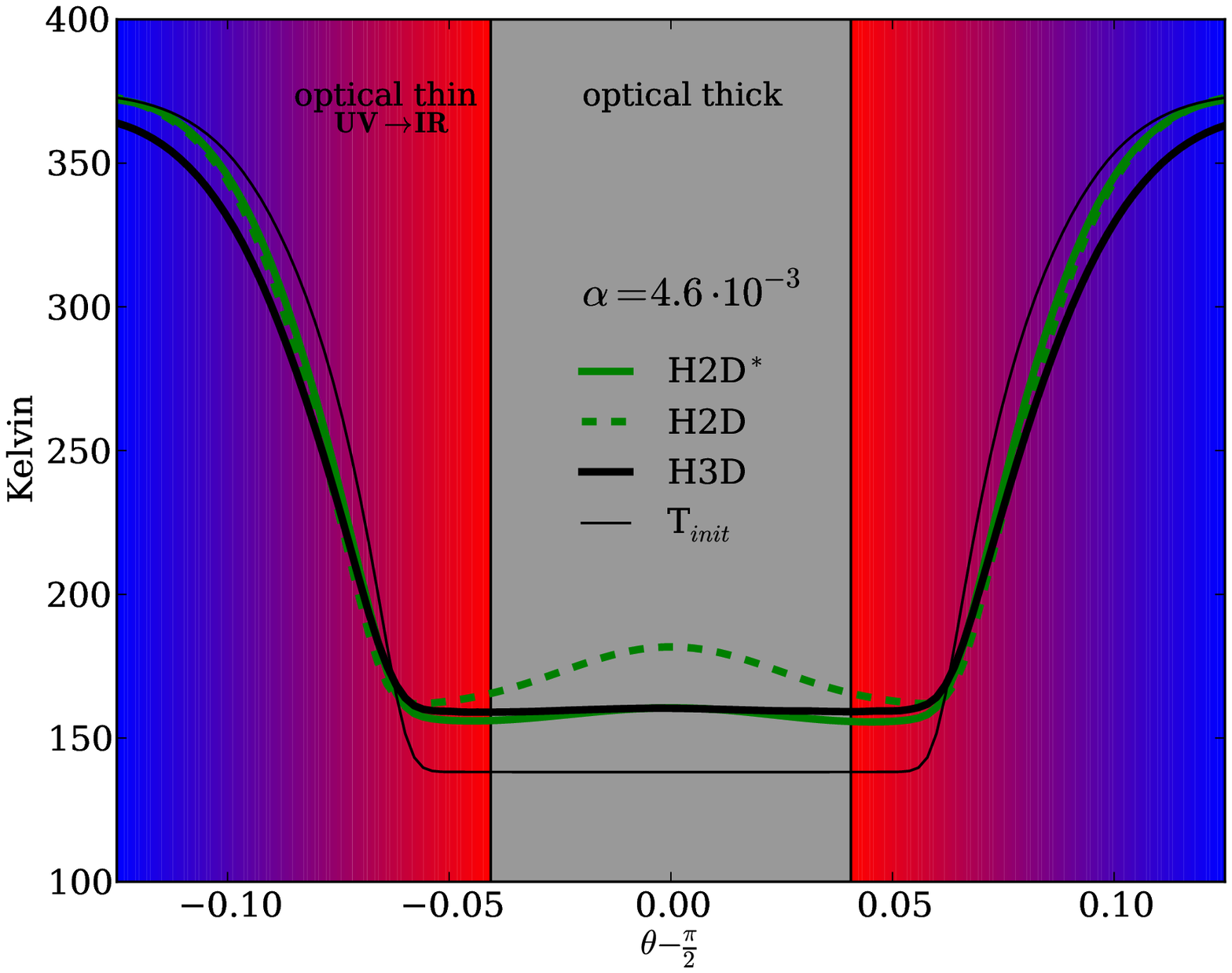,scale=0.40}
\psfig{figure=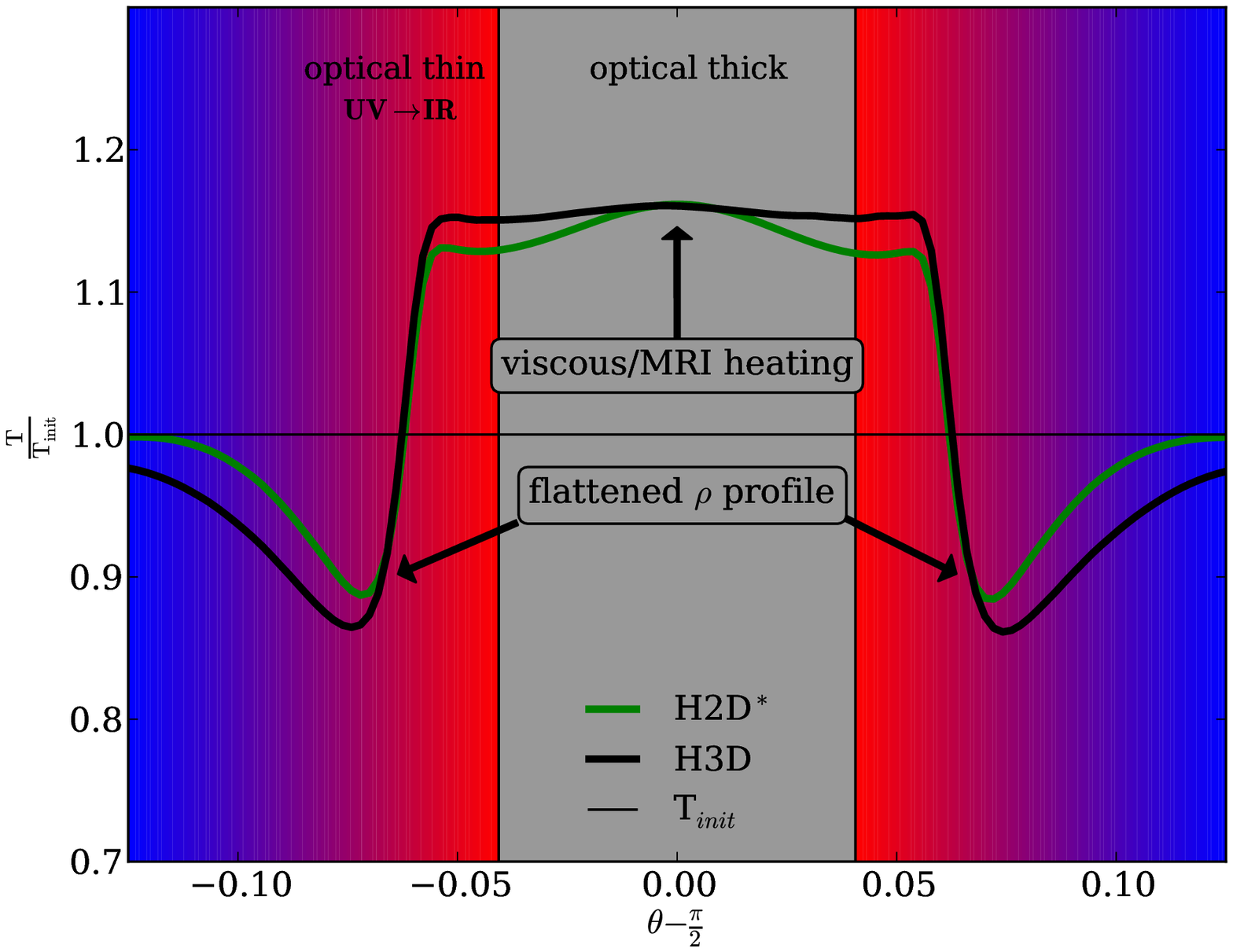,scale=0.40}
\caption{Top: Vertical temperature profile of the RMHD run ({\it black
  line}) and the viscous RHD runs ({\it green lines}) for the high resolution
  models at 1 AU. Model H2D uses a constant alpha. Model H2D$^*$ uses
  a vertical dependence of $\alpha \sim 1/\rho$ which is more similar to the RMHD run. Bottom:
  Relative temperature profile compared to the initial passive disk,
  for models $\rm H2D^*$ and H3D. On both panels, the background
  color shows the region of the disk where the gas is optical thin to
  its own thermal radiation as well as to the irradiation by the
  star. The grey background color shows the region where the gas is
  optical thick to its own radiation. We note that there is also a
  small region in the disk where the gas is optical thin to its own
  radiation but still optical thick for the irradiation by the star. 
} 
\label{fig:TEMP-VRMHD}
\end{figure}
\begin{figure}
\psfig{figure=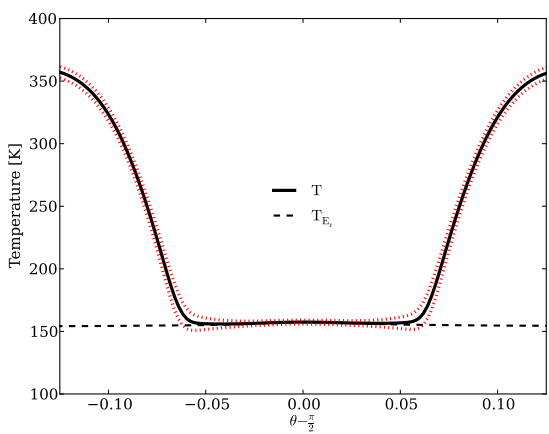,scale=0.40}
\caption{Vertical profile of radiation temperature $\rm
  T_{E_R}=(E_R/a_R)^{1/4}$ ({\it dashed line}) and gas temperature ({\it solid
  line}) line at $1$~AU. We overplot the range of the temperature
  fluctuations ({\it red dotted line}). The peak of relative
  temperature fluctuations $\rm \Delta T/T$ are around 6-7$\%$
  at the $\tau =1$ line ($\theta \approx 0.07$) for the irradiation, compare with
  Fig.~\ref{fig:STR-TIME}} 
\label{fig:TEMP-ERAD}
\end{figure}

 \begin{figure*}
 \psfig{figure=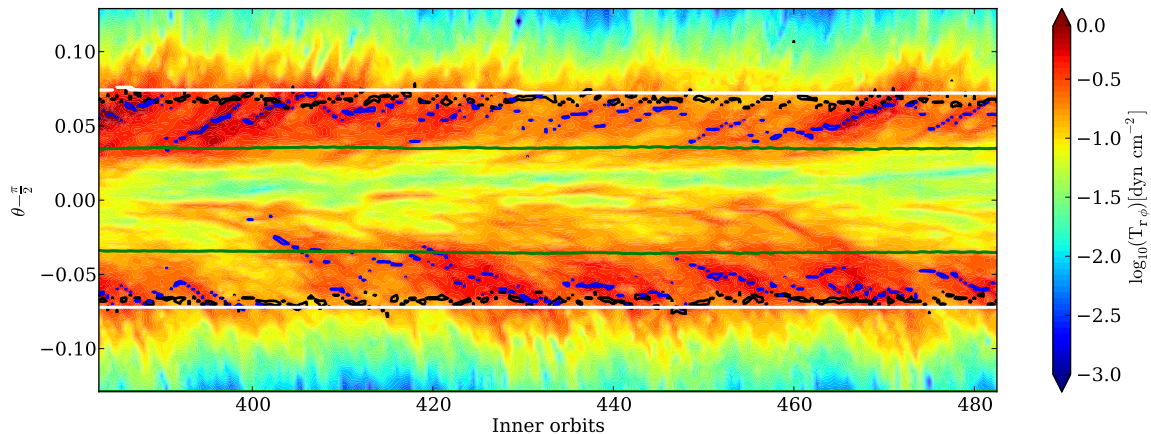,scale=0.50}
 \caption{Time evolution of the azimuthal averaged total stress $\rm
   T_{r\phi}$ over height at 1~AU. We overplot the location of largest
   stress (blue contour), largest relative temperature fluctuations
   (black contour), the vertical location of the $\tau = 1$ position of the
   irradiation (white contour), and the vertical location of the $\tau = 1$ position for
   the local thermal radiation (green contour) for each time bin.} 
 \label{fig:STR-TIME}
 \end{figure*}
%

%
%
The time--averaged vertical temperature profile at $1$~AU in model H3D
is shown in Fig.~\ref{fig:TEMP-VRMHD} and compared with the
temperature vertical profile at the start of the simulation (i.e., when
the disk is in hydrostatic equilibrium). As a consequence of
turbulent heating, the disk midplane temperature increases from around 140~K to
160~K for model H3D. This corresponds to an increase of the disk
pressure scale height of 7 \%. The temperature profile is flat in the
optically thick part of the disk and rising in its upper layers due to the stellar irradiation. At
those locations, we find a small reduction of the temperature by a few
percents (bottom panel of Fig.~\ref{fig:TEMP-VRMHD}). This is because
the disk vertical density profile is flattened in the upper layers as a result of magnetic
support \citep[in agreement with previous results, see for example
][]{hir11}. This shields the disk corona from the incoming irradiation
at a given height compared to the initial model and leads to a small drop
in the temperature. 

We next investigate whether such a vertical temperature profile can be
accounted for in the framework of standard $\rm \alpha$-disk
models. We perform an axisymmetric 2D RHD simulation (model H2D) in the disk poloidal
plane using a constant $\alpha$ viscosity $\rm \nu = \alpha c_s H$ with $\alpha =4.6 \times 10^{-3}$ and the local sound speed $\rm c_s$. The 2D model is
initialized using azimuthally averaged values of the density,
pressure, temperature, azimuthal velocity, and radiation energy
density as obtained in model H3D. The temperature in the 2D viscous
RHD model quickly relaxes into a steady state that is overplotted
on the top panel of Fig.~\ref{fig:TEMP-VRMHD}. As we see, a classical
$\alpha$ disk prescription does not reproduce the correct midplane
temperature. It predicts a midplane temperature of about $\rm T=180$~K, 
higher than that of model H3D and does not display the flat temperature profile
in the optically thick part of the disk. Here, most of the heat is released at
the midplane vicinity due to the scaling of the viscous stress
tensor with density. By contrast, the turbulent stress tensor in our
simulation is rather flat for $|\theta - \pi/2| \le 0.05$ (see
Fig.~\ref{fig:Stress-L}) with variations of only a factor of $\rm \sim
2$. We thus perform an additional 2D RHD simulation, model $\rm
H2D^{*}$, that uses a different prescription for viscosity\footnote{\rm $\rm \nu = \alpha_{mid} \rho_{mid} c_s H$, with $\rm \alpha_{mid}=10^{-3}$ and $\rm \rho_{mid}$ being the midplane density value}, such that
the viscous stress tensor 
remains constant with height below $< \pm 2$
scale heights while it vanishes above that location. 
This $\alpha$ prescription to model the turbulence in one or two dimensional simulation has been used by \citet{kre10,lan13}.
The vertical
temperature profile we obtain in that model is also shown in
Fig.~\ref{fig:TEMP-VRMHD}. It shows midplane temperature and
vertical profile in much better agreement with the full 3D RMHD model
H3D. Besides the gas temperature one can define the radiation
temperature as $\rm T_{E_R}=(E_R/a_R)^{1/4}$. In
Fig.~\ref{fig:TEMP-ERAD} we plot the vertical profile of both
temperatures after 480 inner orbits.  
In the optically thick midplane, the two temperatures are well coupled
due to the high opacity. In the optically thin upper layers the two
temperatures start to diverge due to the irradiation. The radiation
temperature stays at the level of the midplane value.   
Fig.~\ref{fig:TEMP-ERAD} shows also the temperature fluctuations ({\it red
dotted line}). The fluctuations are small.    
The maximum relative temperature fluctuations are close to the $\tau =1$
line of the irradiation with values between $6$ and $7 \%$. In
Fig.~\ref{fig:STR-TIME} we plot the time evolution of the azimuthally averaged total stress $\rm T_{r\phi}$ over height. The blue contour lines show the peak of stress at each time. They follow the butterfly motions of
the mean toroidal field which is triggered by the MRI dynamo
\citep{gre10}. The peak of relative temperature fluctuations are close
to the azimuthally averaged $\tau=1$ absorption layer of the irradiation ({\it white line}). It suggests
that the peak of relative temperature fluctuations is triggered by the 
fluctuations of the $\tau =1$ surface. 

\subsubsection{Heating and cooling rates}
\label{heating_sec}
\begin{figure}
\psfig{figure=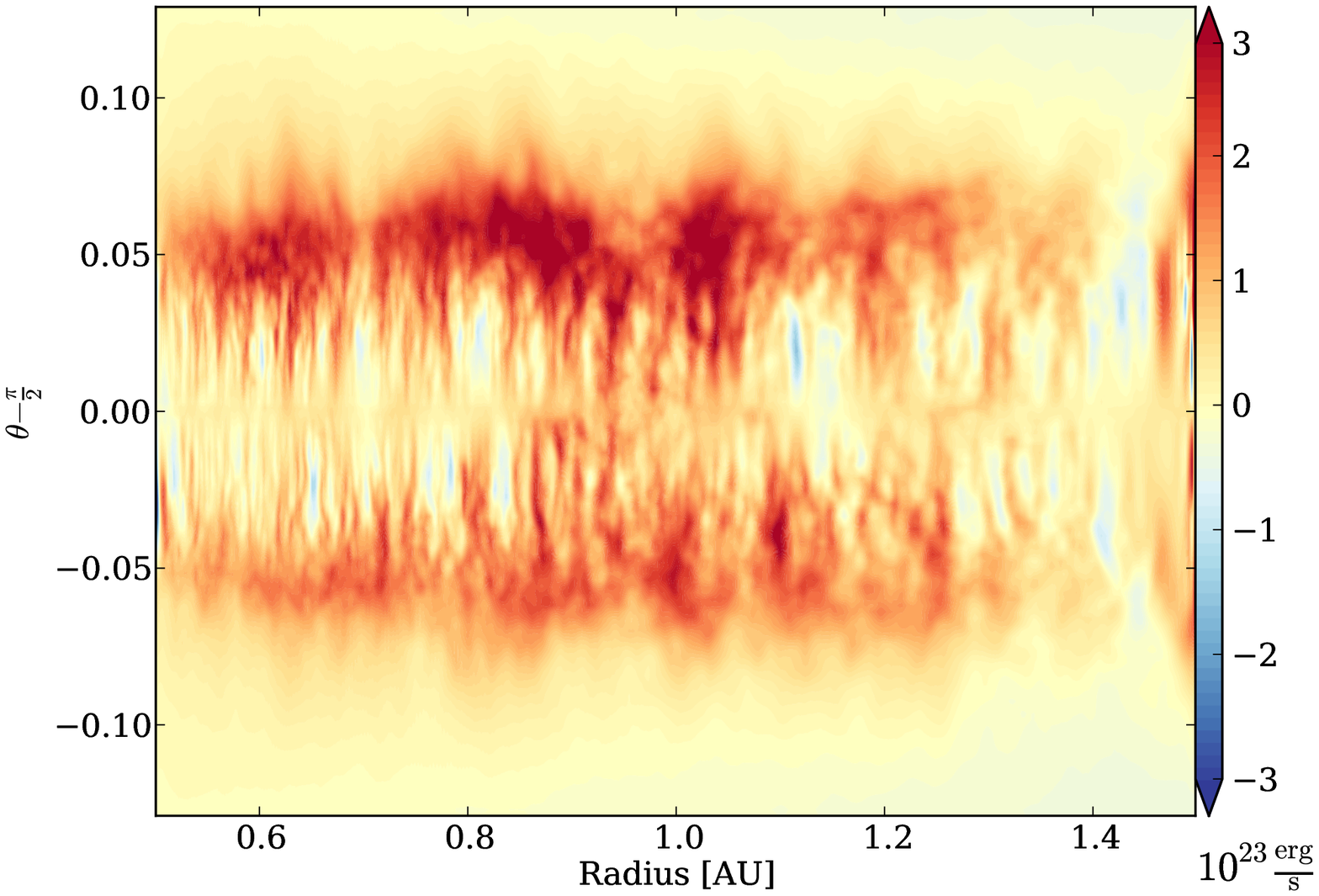,scale=0.40}
\psfig{figure=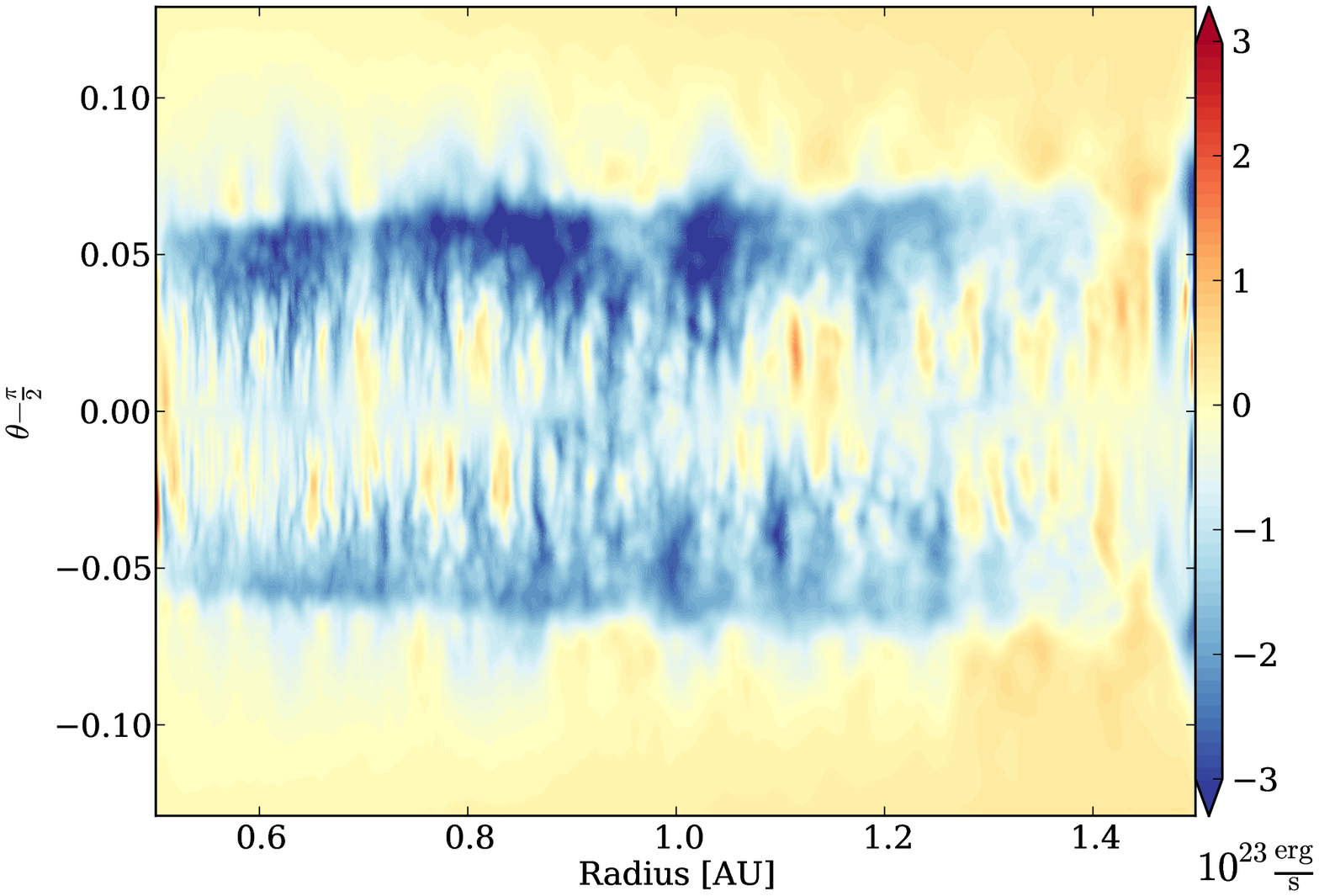,scale=0.40}
\caption{ 2D contour plot of the MHD heating and cooling  $\rm Q^{MHD}$ ({\it left}) and the radiative heating and cooling $\rm Q^{Rad}$ ({\it right}) in units of $10^{23}$, azimuthally and time averaged over 40 inner orbits.}
\label{fig:HEAT-COOL}
\end{figure}
\begin{figure}
\psfig{figure=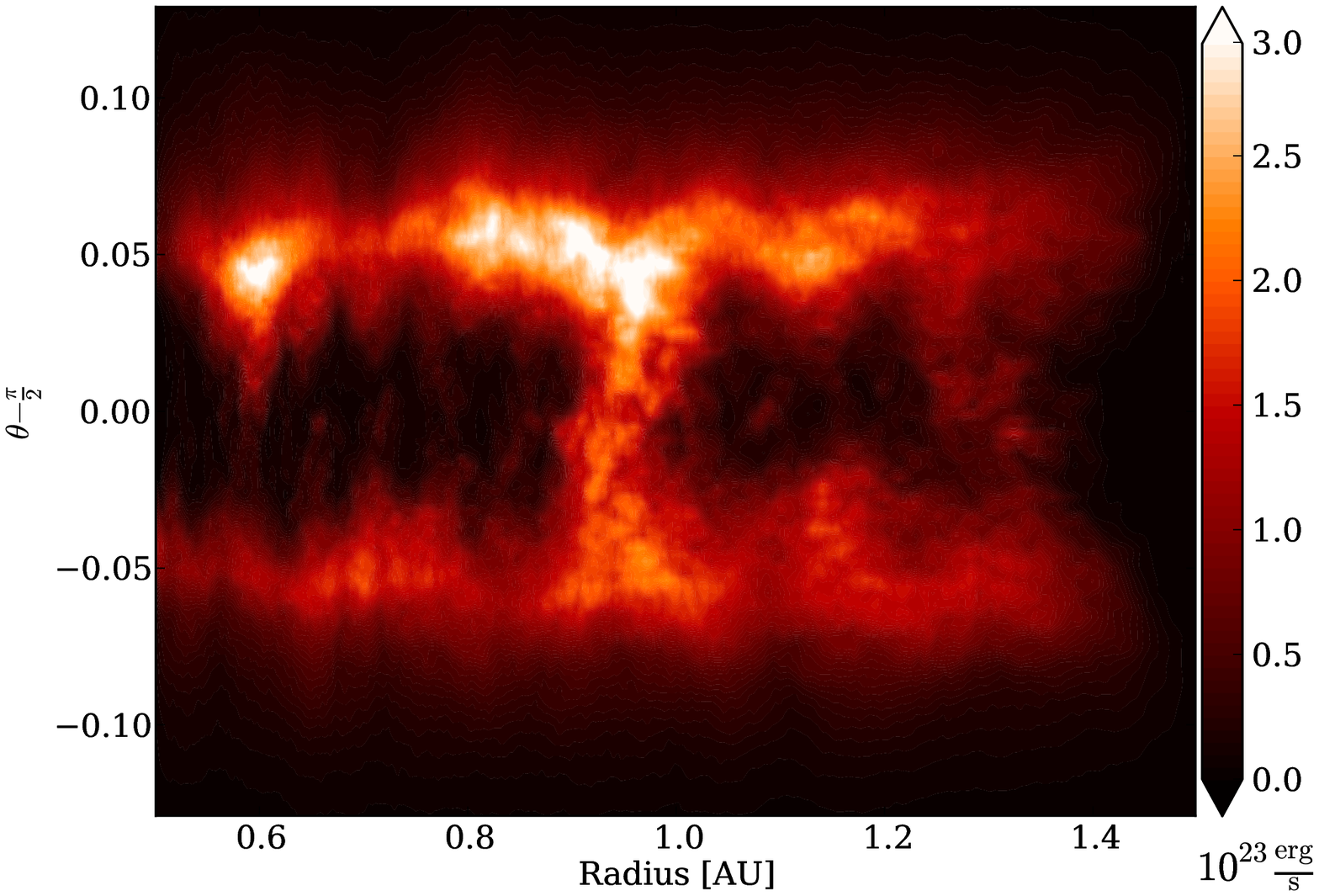,scale=0.40}
\psfig{figure=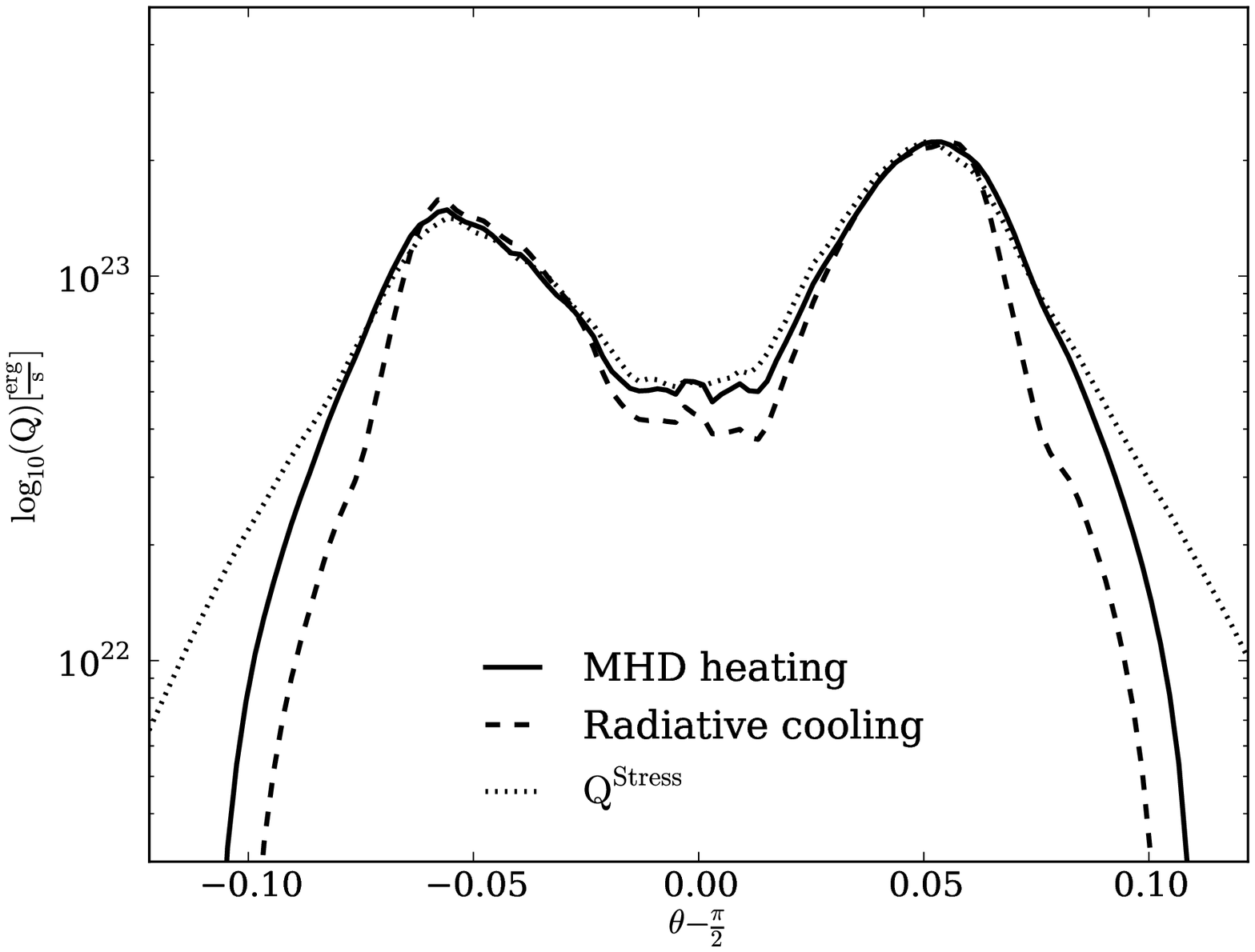,scale=0.40}
\caption{Left: 2D contour plot of expected theoretical heating $\rm Q^{Stress}$ in units of $10^{23}$, azimuthally and time averaged over 40 inner orbits. Right: Vertical profile of the MHD heating $\rm Q^{MHD}$ (solid line), radiative cooling $\rm -Q^{Rad}$ (dashed line) and theoretically expected MHD heating $\rm Q^{Stress}$ (dotted line).}
\label{fig:HEATING}
\end{figure}

In this section we take a closer look at the heating and cooling rates
in the RMHD model H3D. To do so, we proceed as follows: over
any given timestep, we recorded the change of the internal energy $\rm \Delta P_{MHD}/(\Gamma -1)$ that occurred during the MHD step, as well as
the change of internal energy $\rm \Delta P_{rad}/(\Gamma -1)$ that occurred during the
radiative step. The former captures all dynamical heating and cooling
mechanisms, including the advection of energy or the transfers from
kinetic, magnetic or gravitational energy into thermal energy (see 
Eq. 3). We then sum these fractional internal energy changes (divided by the
timestep $\rm \Delta t$ and multiplied by the corresponding cell volume) over a large time interval to compute the
heating and cooling rates associated with dynamical and
radiative processes. These are respectively labelled $\rm Q^{MHD}$ and
$\rm Q^{Rad}$. Fig.~\ref{fig:HEAT-COOL} shows meridional snapshots of
both quantities, respectively on the left ($\rm Q^{MHD}$) and right
($\rm Q^{Rad}$) panels. Both quantities are azimuthally and time
averaged over 40 inner orbits starting after 380 inner orbits. The
plots show that most of 
the disk is being heated with a rate in the order of $\rm 10^{13}
erg/s$ that is mostly released in the disk upper layers, $\rm
\theta-(\pi/2) \sim 0.05$. This corresponds to around two pressure
scale heights above the disk midplane. Radiative cooling (bottom
panel) roughly balances that heating, showing the disk is
approximately in steady state. In order to investigate how much of
that heat can be attributed to turbulent dissipation, we calculate the expected theoretical heating rate $\rm Q^{Stress}$, following \citet{bal99}.
Fig.~\ref{fig:HEATING}, left, shows a meridional snapshot of the turbulent heating
$\rm Q^{Stress}$ that can be computed according to
\begin{equation}
\rm Q^{Stress}=-T_{r\phi} \frac{r \partial \Omega}{\partial
  r} \, .
\end{equation}
The vertical profiles of the heating rates, plotted in Fig.~\ref{fig:HEATING} right, show a good correlation between $\rm Q^{Stress}$ and $\rm Q^{MHD}$. Most of the disk heating can be attributed to MHD
turbulence locally dissipated into heat and only in the upper disk layers, part of the energy is
transported away by waves.

\subsection{Effect of resolution, equation of state and dust--to--gas mass ratio}
\label{varparams_sec}

\begin{figure}
\psfig{figure=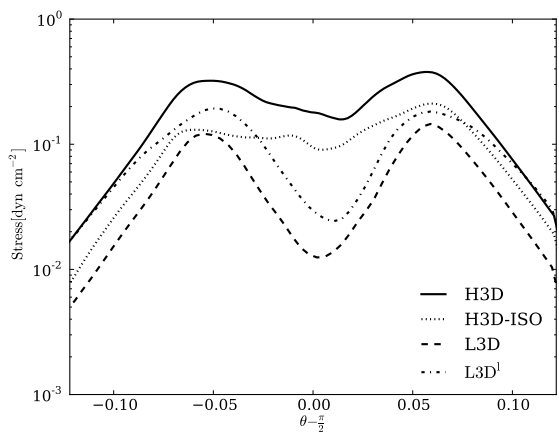,scale=0.40}
\caption{Vertical stress profile in units of dyn.cm$^{-2}$ for the high resolution models H3D ({\it solid line}), H3D-ISO ({\it dotted line}), the low resolution models L3D ({\it dashed line}), and $\rm L3D^l$ ({\it dashed--dotted line}) with reduced amount of dust. }
\label{fig:STRESS-COMP}
\end{figure}

\begin{figure}
\psfig{figure=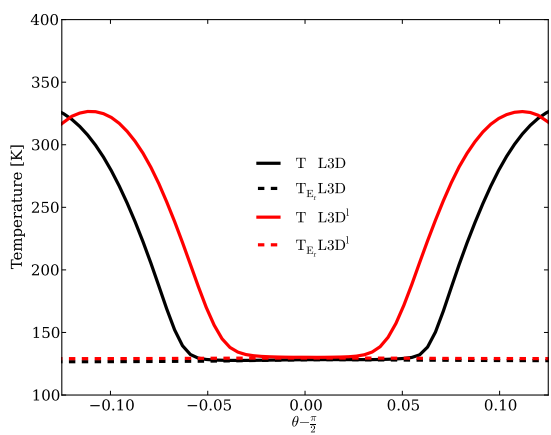,scale=0.40}
\caption{Vertical profile of gas temperature ({\it solid line}) and radiation temperature $\rm T_{E_R}=(E_R/a_R)^{0.25}$ ({\it dashed line}) for models $\rm L3D^l$ ({\it red line}), and $\rm L3D$ ({\it black line}).}
\label{fig:TEMP-COMP}
\end{figure}

\begin{figure}
\psfig{figure=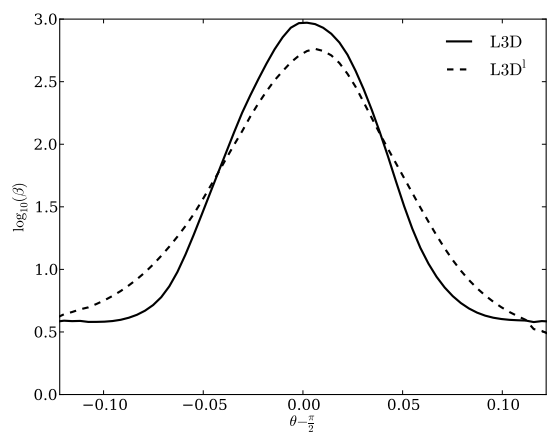,scale=0.40}
\caption{Vertical profile of plasma beta $\rm \beta= 2P/B^2$ for models $\rm L3D$ ({\it solid line}) and $\rm L3D^l$ ({\it dashed line}).}
\label{fig:BETA-COMP}
\end{figure}

We have focused so far on the high resolution model H3D. However, both
the vertical profile of the turbulent stress and the turbulent
velocity depend on several factors of numerical and
physical nature. 

First of all, the spatial resolution of the grid is known to be of 
importance. This is a particularly constraining problem in global
simulations. Recently the convergence and the effect of resolution in
global adiabatic \citep{haw13} and locally isothermal \citep{par13}
simulations were investigated. A convergence study in fully
radiative global simulations is difficult to achieve and would
go beyond the scope of this paper. As a first step in that
direction, we nevertheless present the results of model L3D in which
the resolution is half compared to model H3D. In this low
resolution simulations, there are seven grid cells per pressure scale
height. This is not enough to resolve properly the MRI \citep{flo10,sor12} which leads to
a reduction of the total accretion stress. The normalized total
accretion stress $\alpha$ varies between $4.6 \times 10^{-3}$ for model $\rm
H3D$ and $2.6 \times 10^{-3}$ for model $\rm L3D$. As shown in Fig.~\ref{fig:STRESS-COMP}, the stress vertical profiles of
$\rm T_{r\phi}$ in both models are significantly different. At the
midplane the stress in model L3D drops by one order of magnitude. This is expected as in stratified MRI simulation it becomes more
difficult to resolve properly the MRI at the midplane due to its low magnetization \citep{fro06,flo11}. 

Second, the isothermal model H3D-ISO shows a reduced stress compared to the full RMHD model H3D.
It decreases from $4.6 \times 10^{-3}$ to $2.5 \times 10^{-3}$.
Such a trend of increased turbulence in radiative models was also
suggested by \citet{fla10}. Nevertheless, as shown in
Fig.~\ref{fig:STRESS-COMP}, the vertical profile of the stress has a
similar shape for both models.   

The last effect we want to discuss is the influence of the dust--to--gas mass ratio. In model $\rm L3D^l$, we reduce it by one 
order of magnitude to $10^{-4}$. As shown on Fig.~\ref{fig:RMHD}, this shifts the
irradiated hotter disk region down to the midplane. In Fig.~\ref{fig:TEMP-COMP}, we
plot the temperature profiles at 1~AU, averaged
over azimuth and time between 200 and 400 inner orbits. As the disk
becomes hotter, the sound speed increases and a
higher saturation level of the MRI is expected \citep{bal98}. An effect of increased turbulence can be seen by comparing model $\rm L3D$
({\it dashed line}) and model $\rm L3D^l$ ({\it dashed--dotted line}) in Fig.~\ref{fig:STRESS-COMP}: the
vertical profile of model $\rm L3D^l$, shows an
overall larger stress than model $\rm L3D$. 
The total normalized stress is increased by 25\%, see Table 1. 
But more important is the position of the maximum stress. 
The hotter temperature region has shifted by $\Delta \theta \sim 0.02$ down to the midplane, but the peak stress is still located 
at $\theta-\pi/2 \sim 0.05$. This result indicates again that the position of the maximum stress due to MRI is independent of the vertical temperature profile.
This position seems also independent of resolution by comparing model $\rm L3D$ and model $\rm H3D$.

The position of the maximum of the stress is connected to the plasma beta.
The vertical profiles of $\beta$ for the models $\rm L3D$ and $\rm L3D^l$ are shown in Fig.~\ref{fig:BETA-COMP}, using the same time and space average.
Even if the temperature profiles are quite different, the plasma beta value drops at the same height ($\theta \sim 0.05$) in both
models. All these results indicate that the vertical shape and especially the location of the peak of MRI turbulent magnetic fields are independent 
from the vertical temperature profile.

\section{Conclusion and future work}

In this paper, we successfully implemented in the PLUTO code a FLD
method in spherical coordinates, including frequency dependent
irradiation by a star. It is well
adapted to performing global simulations of irradiated accretion disks
such as protoplanetary disks. The FLD module has serial performances
that are three times faster than the MHD part even for a mainly optically
thin disk setup. We performed the first global 3D
radiation magneto-hydrodynamics simulations of an irradiated and turbulent protoplanetary disk. 
The disk parameters were inspired by that of the system AS 209 in
the star-forming region Ophiuchus \citep{and09} for which there are
strong observational constraints. 
The simulations started from a
radiative hydrostatic disk which becomes MRI unstable, turbulent, and
finally develops into a steady state with typical $\alpha$ 
values of a few times $10^{-3}$, comparable to published simulations
of the same kind that use a locally isothermal equation of state. 
We investigated the turbulent properties and compared the disk structure
with classical viscous disk models. Our findings are:
\begin{itemize}
\item The vertical temperature profile showed no temperature peak at
  the midplane as in classical viscous disk models \citep{dal98}.  
A roughly flat vertical temperature profile established in the disk
optically thick region close to the midplane.  
We reproduced the midplane temperature from the full 3D RMHD run using
2D viscous disk simulations in which the stress tensor is constant in
the bulk of the disk and vanishes in the disk corona. 
A simple prescription is given with the turbulent stress being
constant in the vertical direction within two pressure scale
heights of the midplane, and vanishing above. Such a simple
prescription gives a satisfying account of the results.
\item The main heating in the turbulent disks was dominated by the
  $\rm T_{r\phi}$ stress tensor. We observed a heating of the order of
  $\rm 10^{23}$ erg.s$^{-1}$, mainly released in the disk upper heights. 
\item The temperature fluctuations in the disk were small and of the order of 1\%. A small increase was observed close to the transition region where the disk got heated by the irradiation from the star with fluctuations up to 6\%. 
\item The turbulent magnetic fields reached field strengths of about $1$ to
  $10$ Gauss at the midplane. 
The turbulent velocity of
  the gas was around $10$ to $\rm 100$~m/s at the midplane, and
  up to 1000~$\rm m/s$ in the disk upper heights.  
\end{itemize}   

We want to point out some limitations of our work.
The first one is the distribution and abundance of small sized
  dust particles. Indeed, the latter strongly affects the disk
  temperature vertical profile (see Fig.~\ref{fig:RMHD}). For our
  model we choose the disk AS 209, which has a relative low dust
  abundance compared to other protostellar systems \citep{and09}. For
  the dust surface density we used $\rm 0.017$ g.cm$^{-2}$ at $1$~AU
  for grain sizes $\le 1$~$\mu$m. A much larger amount of small sized
  dust is difficult to include as it needs a much larger vertical
  extent to obtain the optical thin irradiated region. At the same
  time such large extents in stratified MHD turbulent simulation are
  difficult to perform. 
In our simulations we used a fixed dust--to--gas mass ratio.
In contrast, a smaller dust amount over most of the disk height is expected in weakly turbulent disks, e.g. $\rm \alpha^{turb} < 10^{-2}$ \citep{zso11,aki13}. 
All these points show that the total amount and distribution of small dust particles is rather uncertain.
These simulations should be thought of as a proof of concept that RMHD simulations of turbulent protoplanetary disks are now feasible given the current computational resources. 

Another limitation is our use of the ideal MHD approximation. 
It is well known that the electron fraction is 
so low at $1$ AU in protoplanetary disks that dissipation terms (Ohmic
resistivity, ambipolar diffusion, and Hall term) are
important \citep{oku11,bai11,dzy13}. 
These dissipative processes are expected to stabilize the MRI in the bulk of
the disk, producing a laminar dead zone around the disk equatorial
plane \citep{gam96}. 
The presence of a dead
zone and the consequences of the various dissipative 
processes at play will mainly affect the vertical profiles of the
turbulent stress and heating rate \citep{hir11}. 
These are key aspects
of protoplanetary disks dynamics that should be included in future
simulations performed in the planet forming regions of protoplanetary disks.

%

Our current implementation assumes that the gas and dust temperatures are perfectly coupled. Recent models
by \citet{aki13} predict photoelectric heating as a
dominant heating source for the gas affected by UV flux. We
thus expect that the gas temperature and even the dust temperature \citep{aki11} to be higher in the irradiated
regions than presently estimated in our models.
Detailed studies of the flow in the corona (for example reconnection and heating events)
should include this effect to be meaningful. This would be the purpose
of future developments of our numerical scheme.

%
%
%
\section*{Acknowledgements}

We thank Rolf Kuiper for several helpful comments and discussion during this project.  
We thank Andrea Mignone for supporting us with the newest PLUTO code
including the FARGO advection scheme. 
Parallel computations have been performed on the Genci supercomputer
'curie' at the calculation center of CEA TGCC. 
The research leading to these results has received funding from the 
European Research Council under the European Union's Seventh Framework
Programme (FP7/2007-2013) / ERC Grant agreement nr. 258729. 

\appendix

\section{Flux--limited--diffusion method}
\subsection{Numerical scheme}
\label{fld_sec}

We discretize the equations (6-7) in spherical coordinates using a finite
volume formulation and a fully implicit scheme 
\begin{equation}
\begin{aligned}
\rm \frac{c_v^n T_{i,j,k}^{n+1} - c_v^n T_{i,j,k}^{n}}{\Delta t} = -  \kappa_{P,i,j,k}^n(T_{i,j,k}^n) \rho_{i,j,k}^n c (a_R (T_{i,j,k}^{n+1})^4 - E_{R,i,j,k}^{n+1} )\\ \rm + \frac{S^r_{i+1/2} F_{*,i+1/2,j,k} - S^r_{i-1/2} F_{*,i-1/2,j,k}}{V_i^r}
\end{aligned}
\end{equation}
%
%
%
\begin{equation}
\begin{aligned}\rm \frac{E_{R,i,j,k}^{n+1} - E_{R,i,j,k}^n}{\Delta t} -
\rm \left ( \frac{c \lambda}{\kappa_R \rho} \right )_{i+1/2,j,k}^n \frac{S^r_{i+1/2}}{V_i^r} \frac{E_{R,i+1,j,k}^{n+1} - E_{R,i,j,k}^{n+1} }{\Delta r}\\
\rm + \left ( \frac{c \lambda}{\kappa_R \rho} \right )_{i-1/2,j,k}^n  \frac{S^r_{i-1/2}}{V_i^r} \frac{E_{R,i,j,k}^{n+1} - E_{R,i-1,j,k}^{n+1} }{\Delta r}\\ 
\rm - \left (\frac{c \lambda}{\kappa_R \rho} \right )_{i,j+1/2,k}^n  \frac{S^\theta_{j+1/2}}{r V_j^\theta}  \frac{E_{R,i,j+1,k}^{n+1} - E_{R,i,j,k}^{n+1} }{r \Delta \theta}\\ 
\rm + \left (\frac{c \lambda}{\kappa_R \rho} \right )_{i,j-1/2,k}^n \frac{S^\theta_{j-1/2}}{r V_j^\theta}  \frac{E_{R,i,j,k}^{n+1} - E_{R,i,j-1,k}^{n+1} }{r \Delta \theta} \\ 
\rm -  \left (\frac{c \lambda}{\kappa_R \rho} \right )_{i,j,k+1/2}^n  \frac{E_{R,i,j,k+1}^{n+1} - E_{R,i,j,k}^{n+1} }{r^2 \sin{\theta}^2 (\Delta \phi)^2}\\ 
\rm + \left (\frac{c \lambda}{\kappa_R \rho} \right )_{i,j,k-1/2}^n   \frac{E_{R,i,j,k}^{n+1} - E_{R,i,j,k-1}^{n+1} }{r^2 \sin{\theta}^2 (\Delta \phi)^2}\\
\rm = \kappa_{P,i,j,k}(T) \rho c (a_R (T_{i,j,k}^{n+1})^4 - E_{R,i,j,k}^{n+1} )
\end{aligned}
\label{eq:RMHDDISC}
\end{equation}
with the specific heat capacity $\rm c_v=\rho k_B / (\mu_g u (\Gamma -1))$, the geometrical terms $\rm S_i^r = r_i^2$,  $\rm S_j^\theta = |\sin{\theta}|$, $\rm V_i^r=\frac{1}{3}(r_{i+1/2}^3 - r_{i-1/2}^3)$, $\rm V_j^\theta= |\cos{\theta_{j-1/2}} - \cos{\theta_{j+1/2}} | $, and the irradiation flux $\rm F_*$.
Equations A.1 and A.2 are coupled by linearizing the term proportional
to $\rm T^4$ that appears in both equations and neglecting the high order
term \citep{com11} 
\begin{equation}
\rm (T^{n+1})^4 =  4 (T^n)^3 T^{n+1} - 3 (T^n)^4.
\label{eq:LIN}
\end{equation}
This approximation is valid if the change of the temperature $\rm
\Delta T = (T^{n+1} - T^{n})$ is small. The maximum change of the
relative temperature $\rm \Delta T/T$ per time-step during our RMHD disk
simulations is always below $0.01$. 
Using this expression we can combine Equations A.1 and A.2 and
construct the matrix that needs to be inverted.
We note that in this version the scheme is first order in time due to the first-order backward Euler step, used for the time integration.

\subsection{Test in spherical geometry}
\label{test_sec}
\subsubsection{Diffusion test}
%
%
%
%
%
%
\begin{figure}
\hspace{-1.5cm}
\begin{minipage}{0.2\textwidth}
\psfig{figure=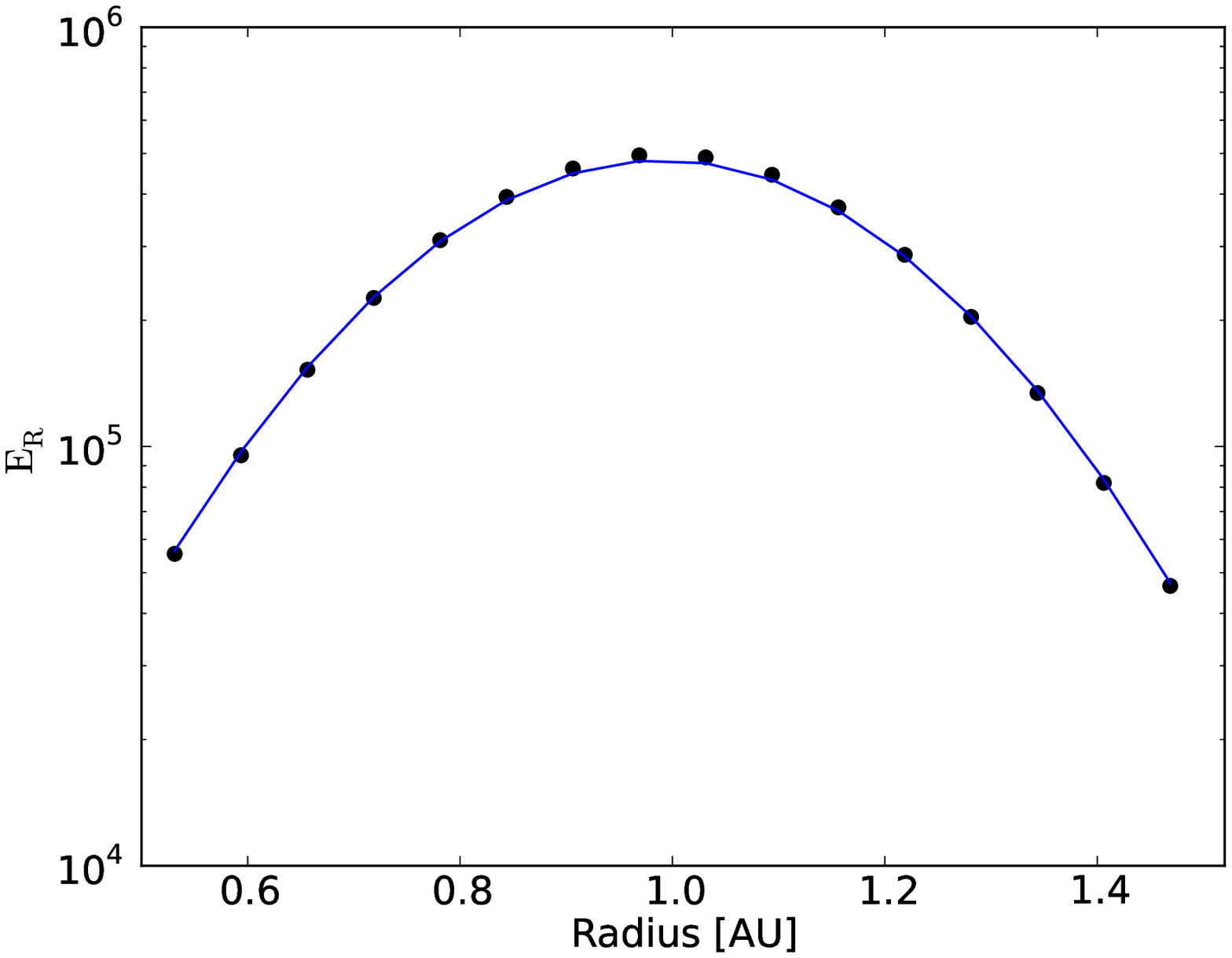,scale=0.24}
\psfig{figure=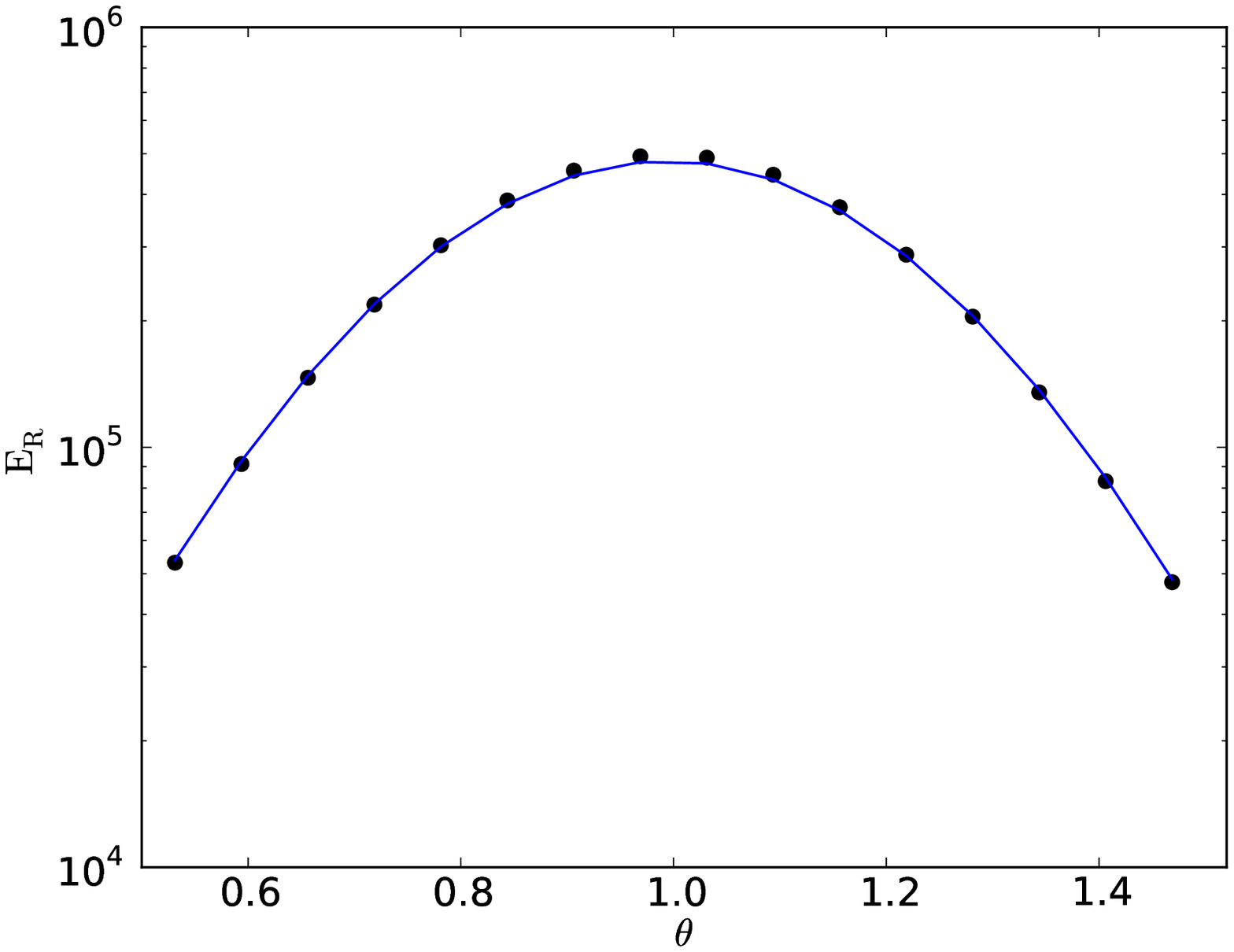,scale=0.24}
\psfig{figure=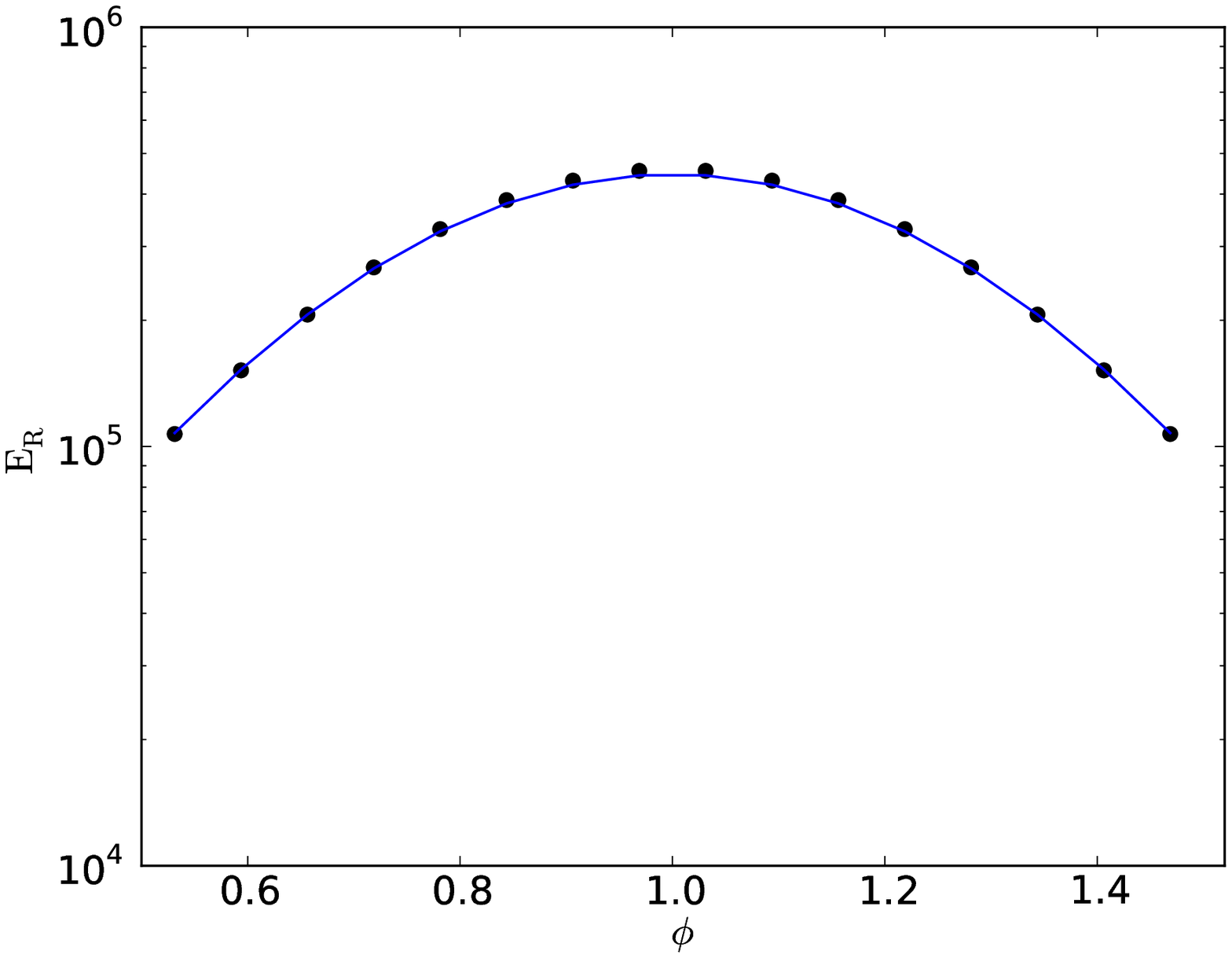,scale=0.24}
\end{minipage}
\hspace{0.8cm}
\begin{minipage}{0.2\textwidth}
\psfig{figure=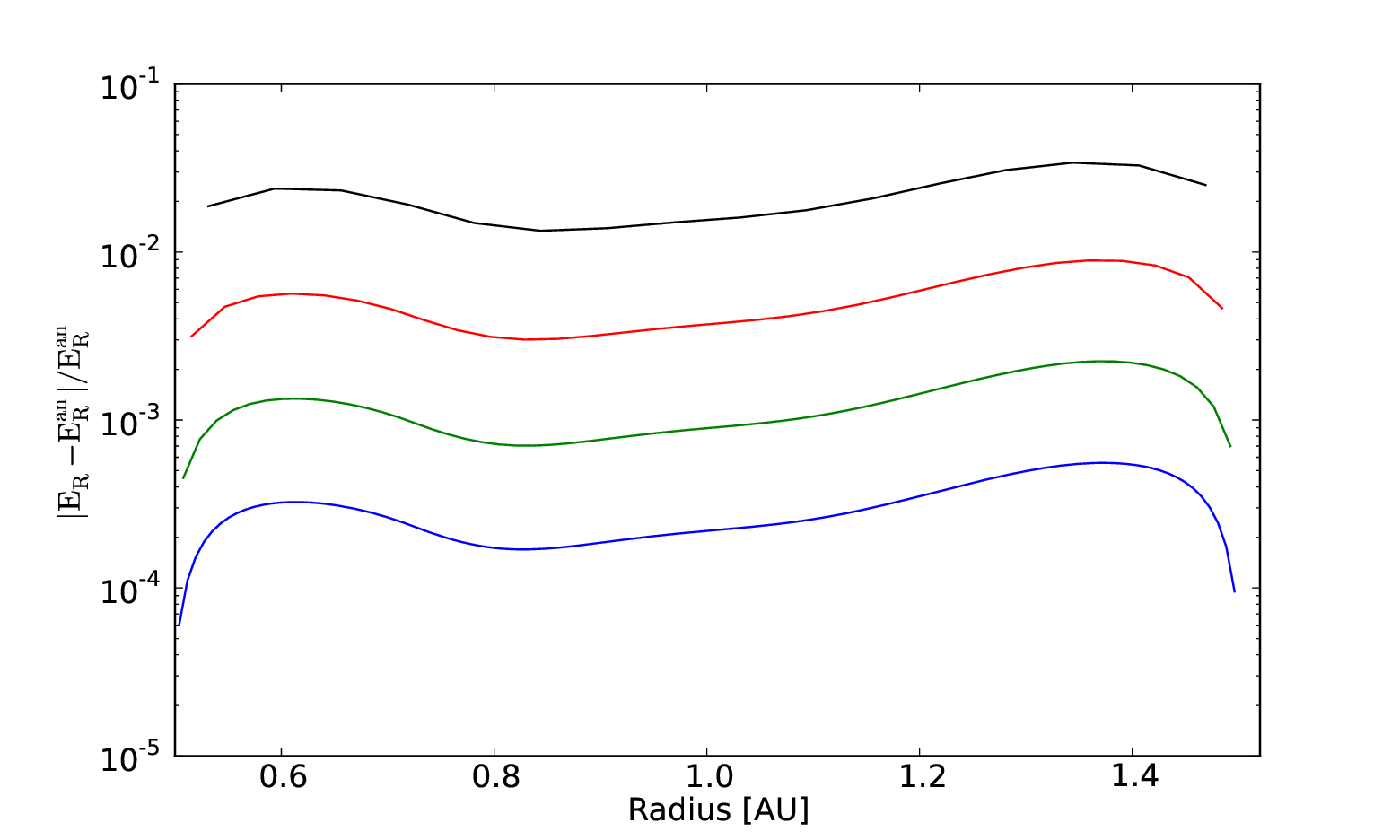,scale=0.24}
\psfig{figure=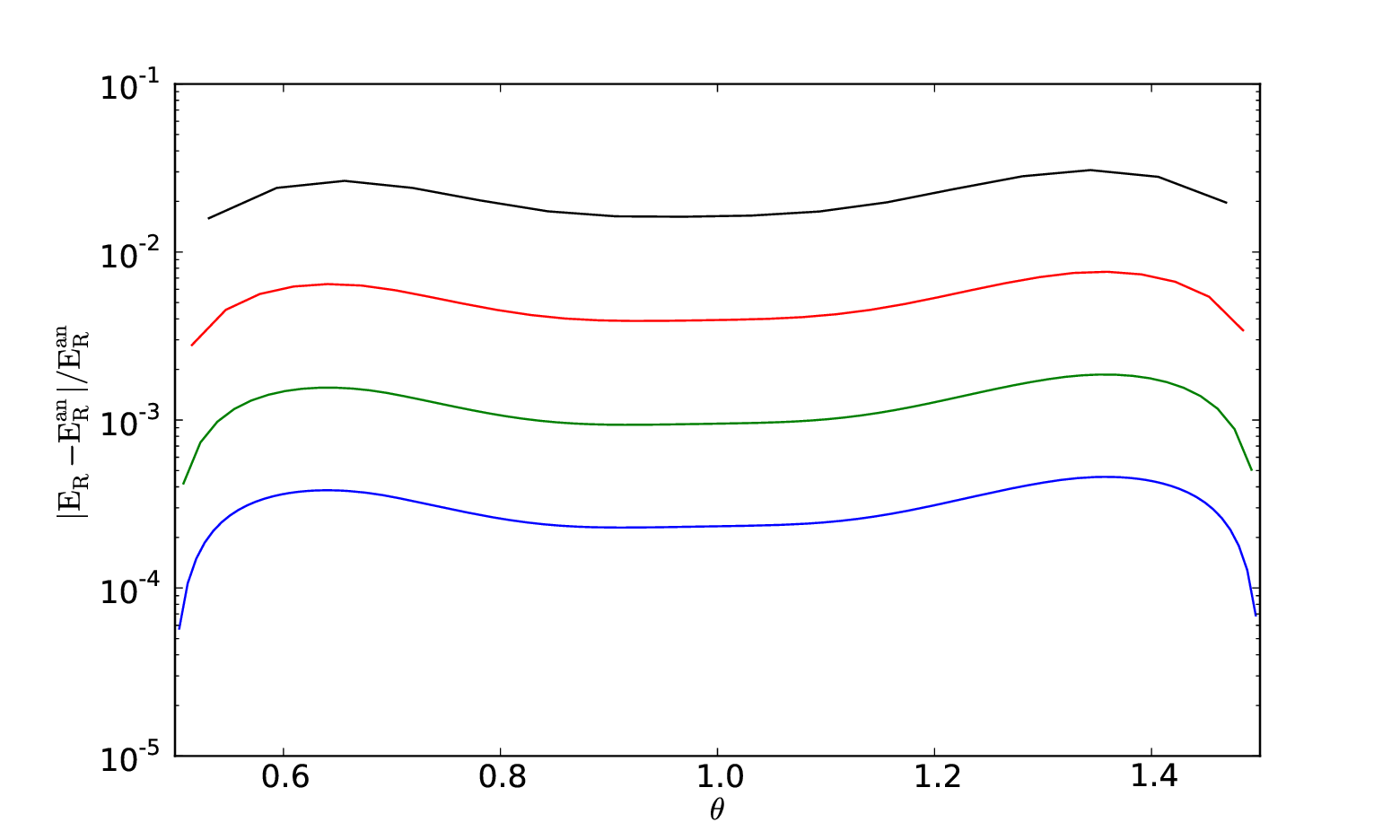,scale=0.24}
\psfig{figure=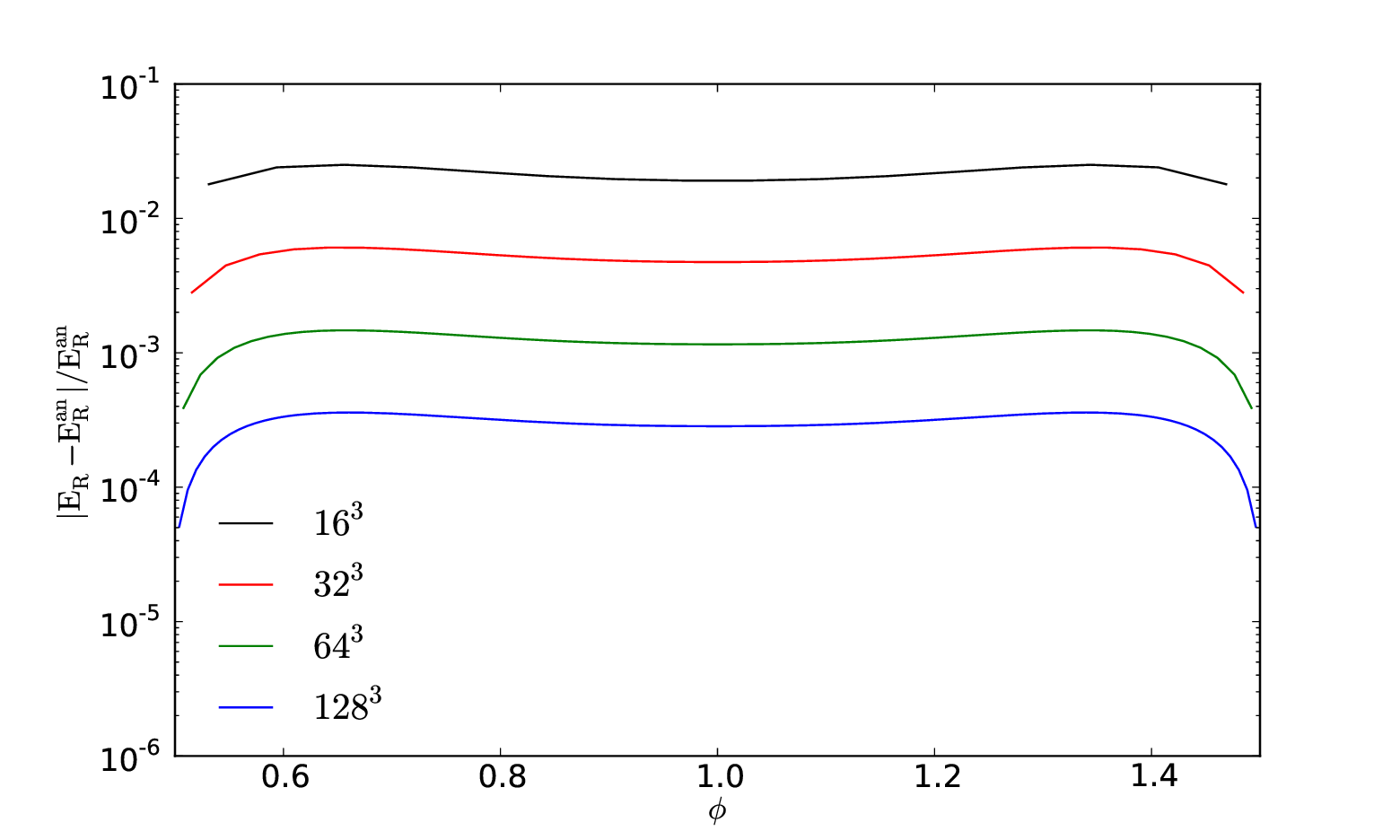,scale=0.24}
\end{minipage}
\caption{Left: Final profiles of $\rm E_R$ ({\it dots}) over radius, $\theta$ and $\phi$ (top to bottom) for the low resolution case. Overplotted is the analytical value ({\it solid line}). Right: Relative error from the analytical profile for the different resolution runs and as a function of radius, $\theta$ and $\phi$.
}
\label{fig:GEO1}
\end{figure}
\begin{figure}
\hspace{-0.5cm}
\begin{minipage}{0.2\textwidth}
\psfig{figure=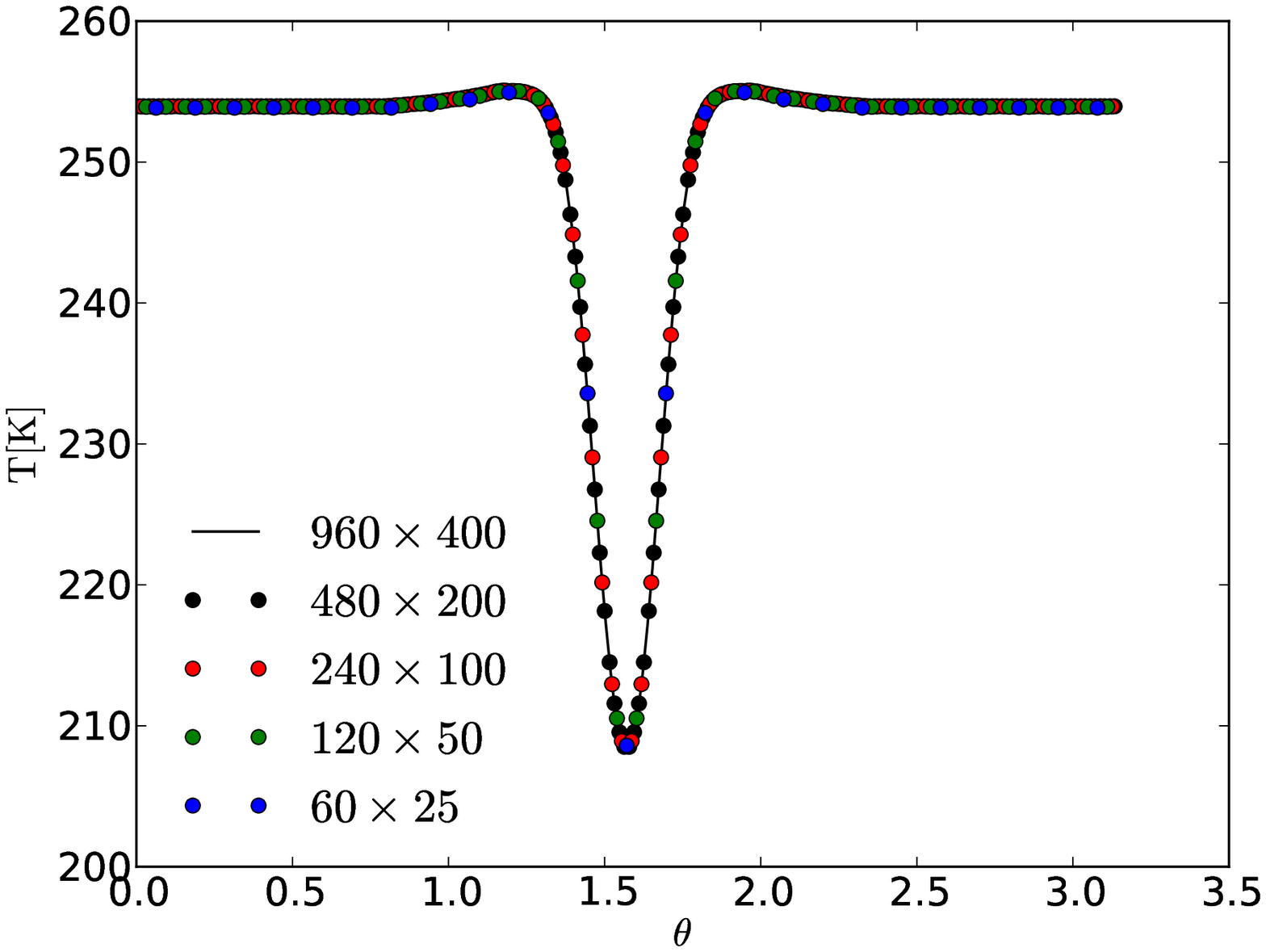,scale=0.24}
\end{minipage}
\hspace{0.8cm}
\begin{minipage}{0.2\textwidth}
\psfig{figure=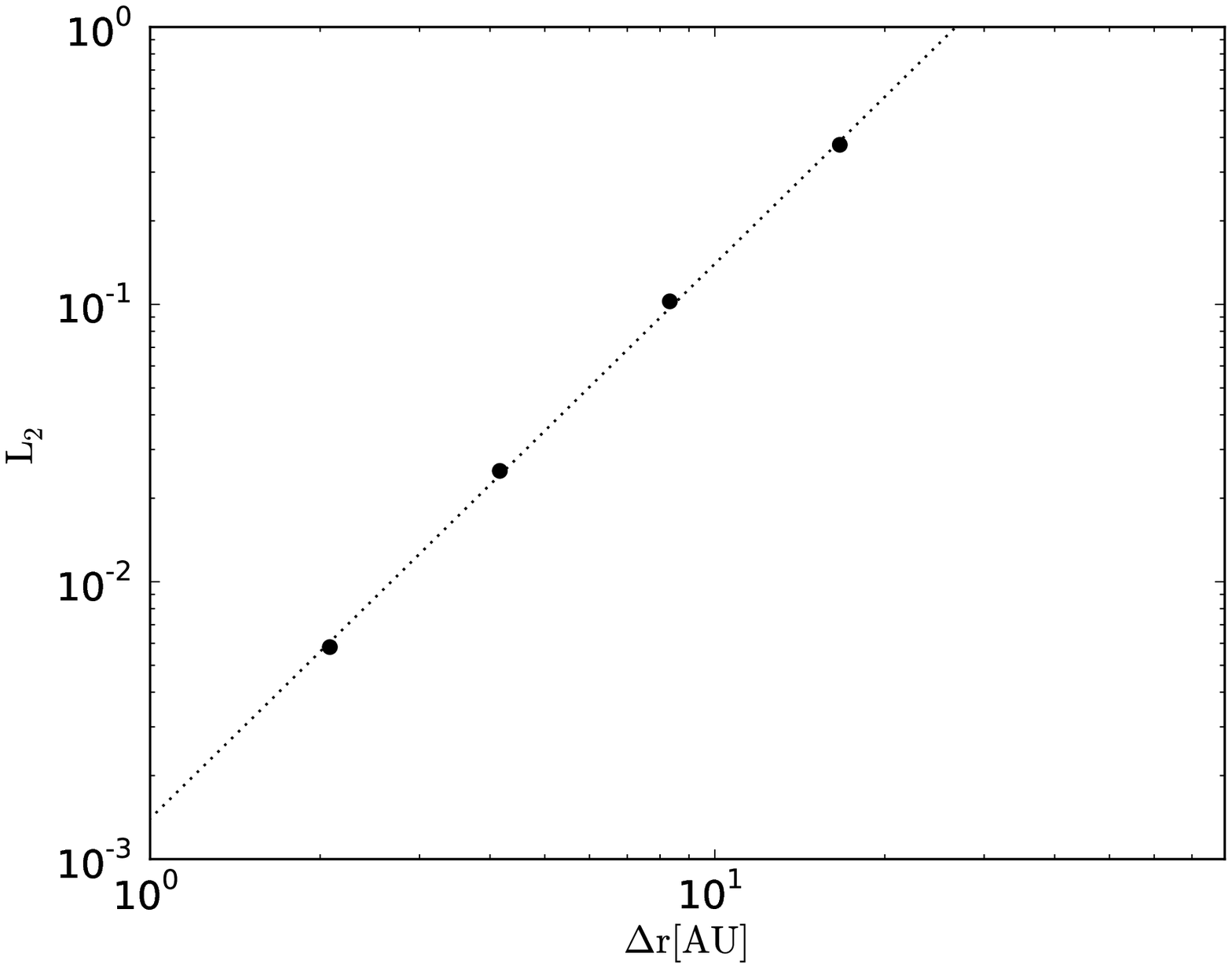,scale=0.24}
\end{minipage}
\caption{Left: Vertical temperature profiles at 2~AU for different resolutions. Right: $\rm L_2$ norm of the relative error ({\it black dots}) over the typical cell size $\rm \Delta r=(r_{out}-r_{in})/N_r$. The dotted line shows the theoretical second order scheme slope $\propto (\Delta r)^2$.}
\label{fig:GEO2}
\end{figure}
In this section we test the diffusion operator in spherical coordinates
\begin{equation}
\rm \frac{\partial E_R}{\partial t} + \nabla \frac{c \lambda}{\kappa \rho} \nabla E_R = 0.
\end{equation}
We set up a domain of $\rm
r:\theta:\phi = (0.5 - 1.5 AU):(0.507 - \pi/2):(0-1)$ with different resolutions
of $16^3$, $32^3$, $64^3$ and $128^3$. 
We use a logarithmic increasing grid in radius so that $\Delta r \sim r \Delta \theta \sim r \Delta \phi$.
The domain is shifted in the $\theta$ direction, to test all the geometrical terms.
As initial conditions we use the Gauss function $$\rm E_R(\vec{x},t_0) =
\frac{10^5\, erg\, cm^{-3}}{(4 \pi D t_0)^{3/2}} e^{-(\vec{x}-\vec{x_0})^2
  / 4 D t_0 } $$ with $\rm D = c/(3 \kappa \rho)$ and the Cartesian position
vector $\vec{x}$. The position $\rm \vec{x_0}$ is placed at $r=1.0$, $\theta=1.0$ and $\phi=1.0$. 
$\kappa$ is fixed to $\rm 1\, cm^2g^{-1}$ and $\rho$ to $\rm 10^{-10}$ g.cm$^{-3}$. We set the flux limiter
$\lambda$ to $1/3$, the initial time to $t_0 = 5000$~s and the boundary conditions to the analytical value for the given time.
We evolve until $t_{final}=55000s$. The timestep for the $16^3$ model is set to $\rm 500s$ which represents a Courant number \footnote{The Courant number for a parabolic problem is defined as $\rm C_p=2\Delta t D/(\Delta x)^2$.} of $0.34$. The timestep is decreased by a factor of 4 each time the resolution is doubled to keep the Courant number constant. Fig.~\ref{fig:GEO1} presents the profile of $\rm E_R$ along each direction for the low resolution case, left column, and the profile of the relative error for each direction and resolution, right column. The low resolution run matches with the analytical profile with relative error around 2\%. We remind that this test problem is difficult to handle in spherical coordinates due to the different grid spacing and especially the change of the volume over radius. As this test problem is time dependent, we observe a first-order convergence rate which is expected, using a first-order time integration \citep{jia12}.
\subsubsection{Full hybrid scheme}
In this second test we repeat the equilibrium setup, see section 2.2, to determine the order of the full hybrid scheme and to verify the equilibrium temperature in the low resolution case. 
We perform a resolution study using five different resolutions $60\times 25$, $120\times 50$, $240\times 100$, $480\times 200$ and $960\times 400$. In this test we set the convergence criteria to $\rm |res|_2/(N_r N_\theta) < 10^{-10}$.
Fig.~\ref{fig:GEO2}, left, shows the vertical temperature profile at 2~AU for the different resolutions. Even the lowest resolution ({\it blue dots}) matches very well with the reference solution ({\it black solid line}). The temperature profile from the highest resolution is taken as reference temperature $\rm T^{ref}$. The order of the scheme can be tested by comparing the $\rm L_2$ norm as a function of the grid spacing. We define the $\rm L_2$ norm as
\begin{equation}
\rm L_2=\sqrt{\rm \frac{1}{N_{cell}} \overset{N_{cell}}{\mathlarger{\sum}} \left[ T^{c} - \frac{ \overset{V^c}{\mathlarger{\sum}} T^{ref}dV^{ref} }{V^c} \right]^2  },
\end{equation}
with the number of cells $\rm N_{cell}$ for a given coarse resolution, the temperature and the corresponding volume of the coarse grid cell $\rm T^c$ and $\rm V^c$, and the temperature and volume of the highest resolution run $\rm T^{ref}$ and $\rm V^{ref}$. We average the reference temperature $\rm T^{ref}$ over the given volume $\rm \overset{V^c}{\mathlarger{\sum}} dV^{ref} = V^{c}$ of the coarse resolution. Using this volume average becomes here important as the method and so the divergence terms are written in the finite volume approach, see Eq. \ref{eq:RMHDDISC}.
In Fig.~\ref{fig:GEO2}, right, 
we show the $\rm L_2$ norm ({\it black dots}) overplotted with the
theoretical slope of a second order scheme. As this test problem is time independent, we are able to obtain second order space accuracy. We note that the irradiation, the heating source, is only radius dependent and so a one dimensional problem.  
%
%
%

\section{Validation of radiative hydrostatic equilibrium}
\label{hydrostatic_sec}
\begin{figure}
\psfig{figure=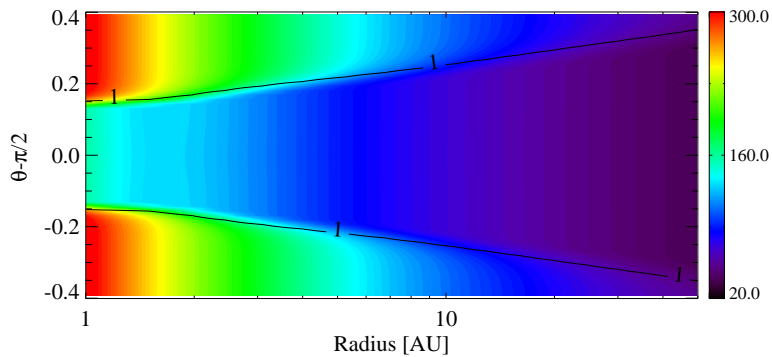,scale=0.50}
\caption{Final 2D temperature distribution in radiation hydrostatic equilibrium. Black solid line shows the $\tau = 1$ line for the irradiation.}
\label{fig:T2D}
\end{figure}

%
%
In this section we test the iterative method presented in
section~\ref{iterative_sec}. To do so, we compare the hydrostatic
disk structure we obtained using that procedure for a given set of disk
parameters with the simple model described by \citet{chi97}. The disk
parameters, chosen to match that work, are as follows: $\rm T_*=4000\,
K$, $\rm M_*=2.5\, M_{\sun}$, $\rm R_*=2.5\, R_{\sun}$, $\rm
\Sigma=1000\, r_{AU}^{-0.5}$ g.cm$^{-2} $, where $\rm r_{AU}$ stands
for the distance to the star in astronomical units. We fix the opacity
to $\rm 1\, cm^2\, g^{-1}$. We use a logarithmically increasing
grid with $384 \times 64$ cells. The radial domain extends from $1$ to
$50$ AU and the poloidal domain covers the range $\theta = \pm 0.4$. 
We follow the iteration procedure presented in section 3.1 to compute
the hydrostatic structure of that disk. The resulting 2D radiative
hydrostatic temperature profile is plotted in Fig.~\ref{fig:T2D}. The
black solid line in Fig.~\ref{fig:T2D} shows the location of the
photosphere.

\begin{figure}
\psfig{figure=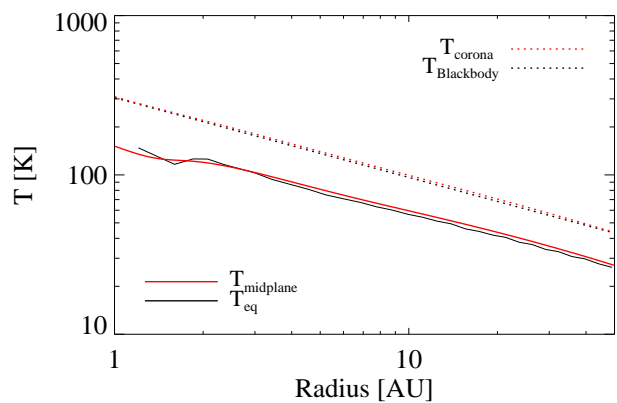,scale=0.60}
\psfig{figure=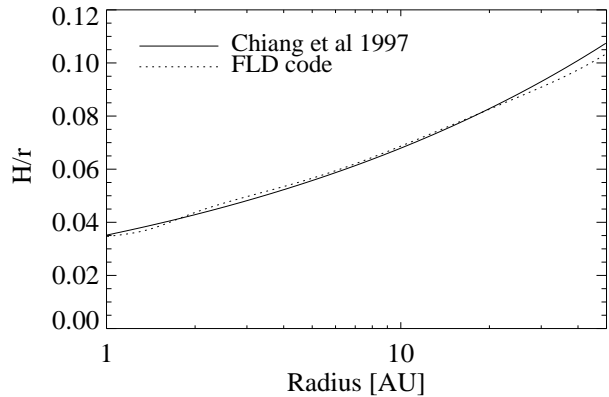,scale=0.60}
\caption{Top: Temperature at the disk photosphere ({\it red dotted line}) and the midplane ({\it red solid line}) overplotted with the analytical prediction ({\it black lines}).
Bottom: Scale height H/r at the midplane ({\it dotted line}), overplotted with the analytical prescription by \citet{chi97} ({\it solid line}).}
\label{fig:CHIANG}
\end{figure}

As mentioned, some of the basic properties of irradiated disks can be
well estimated using the model of \citet{chi97}, the basic physics of
which we review here. The disk is irradiated at the photosphere $\rm 
H_{ph}$. The heating at that location by the incoming irradiation can
be written
\begin{equation}
\rm S_{heat}=\kappa_P \rho \sigma T_{*}^4 \left( \frac{R_*}{r} \right)^2
\rm ds
\label{eq:S_HEAT}
\end{equation}
where $ds=\rm r^2 \sin \theta \Delta \theta \Delta \phi$ is the
irradiated surface element at the disk surface. If we assume isotropic
blackbody cooling at the photosphere we can write the cooling as  
\begin{equation}
\rm S_{cool}=\kappa_p \rho \sigma T^4 (2 r^2 \sin \theta \Delta \theta
\Delta \phi + 2 r \sin \theta \Delta r \Delta \phi) \, .
\label{eq:S_COOL}
\end{equation}
Since irradiation hits the disk surface with a small angle one can
approximate $\rm r \Delta \theta + \Delta r \sim \Delta r$ (in other
words, most of the cooling is done through disk emission in the
vertical direction). Assuming that heating and cooling balance each
other, we find \citep[see also ][ Eq.~(1)]{chi97}
\begin{equation}
\rm T_{eq} = \left (\frac{r \Delta \theta}{2 (r \Delta \theta + \Delta r)}
\rm \right )^{1/4} T_* \left( \frac{R_*}{r} \right)^{1/2} \approx \left(
\rm \frac{r \Delta \theta}{2 \Delta r} \right )^{1/4} T_* \left(
\rm \frac{R_*}{r} \right)^{1/2}. 
\label{eq:T_EQ}
\end{equation}
The term $\rm r \Delta \theta/\Delta r$ is often called the
flaring angle $\rm \alpha^{flare}$ and can be expressed as  
\begin{equation}
\rm \alpha^{flare} = r \frac{\partial}{\partial r} \left(
\frac{H_{ph}}{r} \right) \, .
\label{eq:ALPHA_FLARE}
\end{equation}
Using Eq.~(\ref{eq:ALPHA_FLARE}), we can calculate the flaring angle
of our disk model and derive the expected equilibrium temperature in
our model using Eq.~(\ref{eq:T_EQ}). In Fig.~\ref{fig:CHIANG} (top
panel), we plot the temperature in the corona $\rm T_{corona}$ ({\it
  red dotted line} at $\theta-\pi/2=0.4$)  and in the midplane $\rm
T_{midplane}$ ({\it red solid line}), overplotted with the
corresponding blackbody temperature $\rm T_* (R_*/r
)^{1/2}$ ({\it black dotted line}) and the equilibrium
temperature $\rm T_{eq}$ ({\it black solid line}). The two curves we
obtained using our iterative procedure are in good agreement with the
approximate expressions provided by the \citet{chi97} model. 

In fig.~\ref{fig:CHIANG} (bottom panel), we plot in addition the disk
scale height $\rm H/r$ over radius. When combining
Eq.~(\ref{eq:ALPHA_FLARE}) with the Gaussian vertical profile of the
disk (in the case of an isothermal disk), a simple formulae for its
radial profile is obtained \citep{chi97}: 
\begin{equation}
H/r \propto \left ( \frac{r}{R_*} \right )^{2/7} \, .
\end{equation}
This analytical prediction is confronted on the bottom panel of
fig.~\ref{fig:CHIANG} ({\it solid line}) with our numerical
estimate of the same quantity. Again, the agreement between the two
curves validates our iterative procedure. 
\section{Restarting H3D model}
\label{restart_sec}

As mentioned, we restarted our high resolution model from a magnetic
field configuration after MRI saturation from model L3D. This requires
interpolating the simulation data from a coarse grid to a refined grid
(with refined cells between twice as small as the coarse cells). In
MHD, this is not an immediate procedure if one wants to retain the
solenoidal nature of the magnetic field. In the
constraint transport MHD method, the magnetic fields are located at
cells interfaces. Coarse cells interfaces are refined, and the magnetic
field on those refined surfaces is simply injected from those coarse cell
interfaces. In addition, new interfaces appear in the refined grid that
do not exist in the coarse grid. At those locations, we perform a
linear interpolation of $r^2 \vec{B_r}, \sin{\theta} \vec{B_\theta},
\vec{B_\phi}$ and we use that interpolation to reconstruct the new
magnetic field. This ensures $\nabla \cdot \vec{B} = 0$ on the refined
grid. 

Having the magnetic field at high resolution, we restart the model
first with pure isothermal MHD using the initial density and
temperature profiles derived from hydrostatic equilibrium.  We let the
system relax for around 1000 steps.  
Then we take the velocities and the magnetic fields from the relaxed
state, but the density and temperature profiles again from the
hydrostatic equilibrium. Restarting from this with full radiative RMHD
will result after around 10 outer orbits into a new saturated state
avoiding the linear MRI phase. 
\bibliographystyle{aa}
\bibliography{RMHDDISK}

\begin{thebibliography}{60}
\expandafter\ifx\csname natexlab\endcsname\relax\def\natexlab#1{#1}\fi

\bibitem[{{Akimkin} {et~al.}(2013){Akimkin}, {Zhukovska}, {Wiebe}, {Semenov},
  {Pavlyuchenkov}, {Vasyunin}, {Birnstiel}, \& {Henning}}]{aki13}
{Akimkin}, V., {Zhukovska}, S., {Wiebe}, D., {et~al.} 2013, \apj, 766, 8

\bibitem[{{Akimkin} {et~al.}(2011){Akimkin}, {Pavlyuchenkov}, {Vasyunin},
  {Wiebe}, {Kirsanova}, \& {Henning}}]{aki11}
{Akimkin}, V.~V., {Pavlyuchenkov}, Y.~N., {Vasyunin}, A.~I., {et~al.} 2011,
  \apss, 335, 33

\bibitem[{{Andrews} {et~al.}(2009){Andrews}, {Wilner}, {Hughes}, {Qi}, \&
  {Dullemond}}]{and09}
{Andrews}, S.~M., {Wilner}, D.~J., {Hughes}, A.~M., {Qi}, C., \& {Dullemond},
  C.~P. 2009, \apj, 700, 1502

\bibitem[{{Bai}(2011)}]{bai11}
{Bai}, X.-N. 2011, \apj, 739, 50

\bibitem[{{Balbus} \& {Hawley}(1998)}]{bal98}
{Balbus}, S.~A. \& {Hawley}, J.~F. 1998, Reviews of Modern Physics, 70, 1

\bibitem[{{Balbus} \& {Papaloizou}(1999)}]{bal99}
{Balbus}, S.~A. \& {Papaloizou}, J.~C.~B. 1999, \apj, 521, 650

\bibitem[{{Birnstiel} {et~al.}(2012){Birnstiel}, {Klahr}, \&
  {Ercolano}}]{bir12}
{Birnstiel}, T., {Klahr}, H., \& {Ercolano}, B. 2012, \aap, 539, A148

\bibitem[{{Bitsch} {et~al.}(2013{\natexlab{a}}){Bitsch}, {Boley}, \&
  {Kley}}]{bit13b}
{Bitsch}, B., {Boley}, A., \& {Kley}, W. 2013{\natexlab{a}}, \aap, 550, A52

\bibitem[{{Bitsch} {et~al.}(2013{\natexlab{b}}){Bitsch}, {Crida}, {Morbidelli},
  {Kley}, \& {Dobbs-Dixon}}]{bit13a}
{Bitsch}, B., {Crida}, A., {Morbidelli}, A., {Kley}, W., \& {Dobbs-Dixon}, I.
  2013{\natexlab{b}}, \aap, 549, A124

\bibitem[{{Blaes} {et~al.}(2007){Blaes}, {Hirose}, \& {Krolik}}]{bla07}
{Blaes}, O., {Hirose}, S., \& {Krolik}, J.~H. 2007, \apj, 664, 1057

\bibitem[{{Chiang} \& {Goldreich}(1997)}]{chi97}
{Chiang}, E.~I. \& {Goldreich}, P. 1997, \apj, 490, 368

\bibitem[{{Commer{\c c}on} {et~al.}(2011){Commer{\c c}on}, {Teyssier}, {Audit},
  {Hennebelle}, \& {Chabrier}}]{com11}
{Commer{\c c}on}, B., {Teyssier}, R., {Audit}, E., {Hennebelle}, P., \&
  {Chabrier}, G. 2011, \aap, 529, A35

\bibitem[{{D'Alessio} {et~al.}(1998){D'Alessio}, {Canto}, {Calvet}, \&
  {Lizano}}]{dal98}
{D'Alessio}, P., {Canto}, J., {Calvet}, N., \& {Lizano}, S. 1998, \apj, 500,
  411

\bibitem[{{Decampli} {et~al.}(1978){Decampli}, {Cameron}, {Bodenheimer}, \&
  {Black}}]{dec78}
{Decampli}, W.~M., {Cameron}, A.~G.~W., {Bodenheimer}, P., \& {Black}, D.~C.
  1978, \apj, 223, 854

\bibitem[{{Dittrich} {et~al.}(2013){Dittrich}, {Klahr}, \& {Johansen}}]{dit13}
{Dittrich}, K., {Klahr}, H., \& {Johansen}, A. 2013, \apj, 763, 117

\bibitem[{{Draine} \& {Lee}(1984)}]{dra84}
{Draine}, B.~T. \& {Lee}, H.~M. 1984, \apj, 285, 89

\bibitem[{{Dullemond}(2012)}]{dul12}
{Dullemond}, C.~P. 2012, {RADMC-3D: A multi-purpose radiative transfer tool},
  astrophysics Source Code Library

\bibitem[{{Dullemond} {et~al.}(2002){Dullemond}, {van Zadelhoff}, \&
  {Natta}}]{dul02b}
{Dullemond}, C.~P., {van Zadelhoff}, G.~J., \& {Natta}, A. 2002, \aap, 389, 464

\bibitem[{{Dzyurkevich} {et~al.}(2010){Dzyurkevich}, {Flock}, {Turner},
  {Klahr}, \& {Henning}}]{dzy10}
{Dzyurkevich}, N., {Flock}, M., {Turner}, N.~J., {Klahr}, H., \& {Henning}, T.
  2010, \aap, 515, A70

\bibitem[{{Dzyurkevich} {et~al.}(2013){Dzyurkevich}, {Turner}, {Henning}, \&
  {Kley}}]{dzy13}
{Dzyurkevich}, N., {Turner}, N.~J., {Henning}, T., \& {Kley}, W. 2013, \apj,
  765, 114

\bibitem[{{Flaig} {et~al.}(2009){Flaig}, {Kissmann}, \& {Kley}}]{fla09}
{Flaig}, M., {Kissmann}, R., \& {Kley}, W. 2009, \mnras, 282

\bibitem[{{Flaig} {et~al.}(2010){Flaig}, {Kley}, \& {Kissmann}}]{fla10}
{Flaig}, M., {Kley}, W., \& {Kissmann}, R. 2010, \mnras, 409, 1297

\bibitem[{{Flock} {et~al.}(2010){Flock}, {Dzyurkevich}, {Klahr}, \&
  {Mignone}}]{flo10}
{Flock}, M., {Dzyurkevich}, N., {Klahr}, H., \& {Mignone}, A. 2010, \aap, 516,
  A26

\bibitem[{{Flock} {et~al.}(2011){Flock}, {Dzyurkevich}, {Klahr}, {Turner}, \&
  {Henning}}]{flo11}
{Flock}, M., {Dzyurkevich}, N., {Klahr}, H., {Turner}, N.~J., \& {Henning}, T.
  2011, \apj, 735, 122

\bibitem[{{Fromang} \& {Nelson}(2006)}]{fro06}
{Fromang}, S. \& {Nelson}, R.~P. 2006, \aap, 457, 343

\bibitem[{{Furlan} {et~al.}(2009){Furlan}, {Watson}, {McClure}, {Manoj},
  {Espaillat}, {D'Alessio}, {Calvet}, {Kim}, {Sargent}, {Forrest}, \&
  {Hartmann}}]{fur09}
{Furlan}, E., {Watson}, D.~M., {McClure}, M.~K., {et~al.} 2009, \apj, 703, 1964

\bibitem[{{Gammie}(1996)}]{gam96}
{Gammie}, C.~F. 1996, \apj, 457, 355

\bibitem[{{Gardiner} \& {Stone}(2005)}]{gar05}
{Gardiner}, T.~A. \& {Stone}, J.~M. 2005, Journal of Computational Physics,
  205, 509

\bibitem[{{Gonz{\'a}lez} {et~al.}(2007){Gonz{\'a}lez}, {Audit}, \&
  {Huynh}}]{gon07}
{Gonz{\'a}lez}, M., {Audit}, E., \& {Huynh}, P. 2007, \aap, 464, 429

\bibitem[{{Gressel}(2010)}]{gre10}
{Gressel}, O. 2010, \mnras, 404

\bibitem[{{G{\"u}nther}(2013)}]{gun13}
{G{\"u}nther}, H.~M. 2013, Astronomische Nachrichten, 334, 67

\bibitem[{{Hawley} {et~al.}(2013){Hawley}, {Richers}, {Guan}, \&
  {Krolik}}]{haw13}
{Hawley}, J.~F., {Richers}, S.~A., {Guan}, X., \& {Krolik}, J.~H. 2013, \apj,
  772, 102

\bibitem[{{Hirose} {et~al.}(2006){Hirose}, {Krolik}, \& {Stone}}]{hir06}
{Hirose}, S., {Krolik}, J.~H., \& {Stone}, J.~M. 2006, \apj, 640, 901

\bibitem[{{Hirose} \& {Turner}(2011)}]{hir11}
{Hirose}, S. \& {Turner}, N.~J. 2011, \apjl, 732, L30

\bibitem[{{Jiang} {et~al.}(2012){Jiang}, {Stone}, \& {Davis}}]{jia12}
{Jiang}, Y.-F., {Stone}, J.~M., \& {Davis}, S.~W. 2012, \apjs, 199, 14

\bibitem[{{Johansen} {et~al.}(2009){Johansen}, {Youdin}, \& {Klahr}}]{joh09}
{Johansen}, A., {Youdin}, A., \& {Klahr}, H. 2009, \apj, 697, 1269

\bibitem[{{Koerner} \& {Sargent}(1995)}]{koe95}
{Koerner}, D.~W. \& {Sargent}, A.~I. 1995, \aj, 109, 2138

\bibitem[{{Kolb} {et~al.}(2013){Kolb}, {Stute}, {Kley}, \& {Mignone}}]{kol13}
{Kolb}, S.~M., {Stute}, M., {Kley}, W., \& {Mignone}, A. 2013, ArXiv e-prints

\bibitem[{{Kretke} \& {Lin}(2010)}]{kre10}
{Kretke}, K.~A. \& {Lin}, D.~N.~C. 2010, \apj, 721, 1585

\bibitem[{{Krolik} {et~al.}(2007){Krolik}, {Hirose}, \& {Blaes}}]{kro07}
{Krolik}, J.~H., {Hirose}, S., \& {Blaes}, O. 2007, \apj, 664, 1045

\bibitem[{{Krumholz} {et~al.}(2007){Krumholz}, {Klein}, {McKee}, \&
  {Bolstad}}]{kru07}
{Krumholz}, M.~R., {Klein}, R.~I., {McKee}, C.~F., \& {Bolstad}, J. 2007, \apj,
  667, 626

\bibitem[{{Kuiper} {et~al.}(2010){Kuiper}, {Klahr}, {Dullemond}, {Kley}, \&
  {Henning}}]{kui10}
{Kuiper}, R., {Klahr}, H., {Dullemond}, C., {Kley}, W., \& {Henning}, T. 2010,
  \aap, 511, A81

\bibitem[{{Kuiper} \& {Klessen}(2013)}]{kui13}
{Kuiper}, R. \& {Klessen}, R.~S. 2013, \aap, 555, A7

\bibitem[{{Landry} {et~al.}(2013){Landry}, {Dodson-Robinson}, {Turner}, \&
  {Abram}}]{lan13}
{Landry}, R., {Dodson-Robinson}, S.~E., {Turner}, N.~J., \& {Abram}, G. 2013,
  \apj, 771, 80

\bibitem[{{Levermore} \& {Pomraning}(1981)}]{lev81}
{Levermore}, C.~D. \& {Pomraning}, G.~C. 1981, \apj, 248, 321

\bibitem[{{Masset}(2000)}]{mas00}
{Masset}, F. 2000, \aaps, 141, 165

\bibitem[{{Mignone}(2009)}]{mig09}
{Mignone}, A. 2009, Nuovo Cimento C Geophysics Space Physics C, 32, 37

\bibitem[{{Mignone} {et~al.}(2007){Mignone}, {Bodo}, {Massaglia}, {Matsakos},
  {Tesileanu}, {Zanni}, \& {Ferrari}}]{mig07}
{Mignone}, A., {Bodo}, G., {Massaglia}, S., {et~al.} 2007, \apjs, 170, 228

\bibitem[{{Mignone} {et~al.}(2012){Mignone}, {Flock}, {Stute}, {Kolb}, \&
  {Muscianisi}}]{mig12}
{Mignone}, A., {Flock}, M., {Stute}, M., {Kolb}, S.~M., \& {Muscianisi}, G.
  2012, \aap, 545, A152

\bibitem[{{Miyoshi} \& {Kusano}(2005)}]{miy05}
{Miyoshi}, T. \& {Kusano}, K. 2005, Journal of Computational Physics, 208, 315

\bibitem[{{Okuzumi} \& {Hirose}(2011)}]{oku11}
{Okuzumi}, S. \& {Hirose}, S. 2011, \apj, 742, 65

\bibitem[{{Parkin} \& {Bicknell}(2013)}]{par13}
{Parkin}, E.~R. \& {Bicknell}, G.~V. 2013, ArXiv e-prints

\bibitem[{{Pascucci} {et~al.}(2004){Pascucci}, {Wolf}, {Steinacker},
  {Dullemond}, {Henning}, {Niccolini}, {Woitke}, \& {Lopez}}]{pas04}
{Pascucci}, I., {Wolf}, S., {Steinacker}, J., {et~al.} 2004, \aap, 417, 793

\bibitem[{{P{\'e}rez} {et~al.}(2012){P{\'e}rez}, {Carpenter}, {Chandler},
  {Isella}, {Andrews}, {Ricci}, {Calvet}, {Corder}, {Deller}, {Dullemond},
  {Greaves}, {Harris}, {Henning}, {Kwon}, {Lazio}, {Linz}, {Mundy}, {Sargent},
  {Storm}, {Testi}, \& {Wilner}}]{per12}
{P{\'e}rez}, L.~M., {Carpenter}, J.~M., {Chandler}, C.~J., {et~al.} 2012,
  \apjl, 760, L17

\bibitem[{{Simon} {et~al.}(2012){Simon}, {Beckwith}, \& {Armitage}}]{sim12}
{Simon}, J.~B., {Beckwith}, K., \& {Armitage}, P.~J. 2012, \mnras, 422, 2685

\bibitem[{{Sorathia} {et~al.}(2012){Sorathia}, {Reynolds}, {Stone}, \&
  {Beckwith}}]{sor12}
{Sorathia}, K.~A., {Reynolds}, C.~S., {Stone}, J.~M., \& {Beckwith}, K. 2012,
  \apj, 749, 189

\bibitem[{{Turner}(2004)}]{tur04}
{Turner}, N.~J. 2004, \apjl, 605, L45

\bibitem[{{Turner} {et~al.}(2003){Turner}, {Stone}, {Krolik}, \&
  {Sano}}]{tur03}
{Turner}, N.~J., {Stone}, J.~M., {Krolik}, J.~H., \& {Sano}, T. 2003, \apj,
  593, 992

\bibitem[{{Van der Vorst}(1992)}]{van92}
{Van der Vorst}, H.~A. 1992, SIAM J. Sci. and Stat. Comput, 13, 631–644

\bibitem[{{Zsom} {et~al.}(2011){Zsom}, {Ormel}, {Dullemond}, \&
  {Henning}}]{zso11}
{Zsom}, A., {Ormel}, C.~W., {Dullemond}, C.~P., \& {Henning}, T. 2011, \aap,
  534, A73

\end{thebibliography}
\end{document}